\documentclass{mn2e}  
\usepackage{psfig,lscape}

\def\om{\Omega_p}
\def\len{a_B}
\def\lag{D_L}
\def\vpd{{\cal R}}
\def\kin{{\cal V}}
\def\pin{{\cal X}}
   
\def\hi{H~{\small I}}   
\def\gtsim{\mathrel{\spose{\lower.5ex \hbox{$\mathchar"218$}}
     \raise.4ex\hbox{$\mathchar"13E$}}}
\def\ltsim{\mathrel{\spose{\lower.5ex\hbox{$\mathchar"218$}}
     \raise.4ex\hbox{$\mathchar"13C$}}}

\def\degrees{^\circ}

\def\kms{$\mathrm km\;s^{-1}$}
\def\kmsa{$\mathrm km\;s^{-1}\,arcsec^{-1}$}

\def\mas{mag arcsec$^{-2}$}   
\def\asecpix{arcsec pixel$^{-1}$}  

\def\iraf{{\tt IRAF}}

\def\etal{{et al.}}
\def\eg{{\it e.g.}}

\def\ie{{\it i.e.}}

\def\spose#1{\hbox to 0pt{#1\hss}}

\begin{document}   
   
\title[Measurement of fast bars in a sample of early-type 
barred galaxies]{Measurement of fast bars in a sample of early-type 
barred galaxies}

\author[J.A.L. Aguerri, V.P. Debattista and E.M. Corsini] 
{J. A. L. Aguerri$^{1,2}$\thanks{email: {\tt jalfonso@ll.iac.es}},
Victor P. Debattista$^2$,
and Enrico Maria Corsini$^3$\\
$^1$ Instituto de Astrof\'{\i}sica de Canarias, V\'{\i}a L\'actea s/n.
 E-38200 La Laguna, Spain \\ 
$^2$ Astronomisches Institut, Universit\"at Basel, Venusstrasse 7,
CH-4102 Binningen, Switzerland\\ 
$^3$ Dipartimento di Astronomia,
Universit\`a di Padova, vicolo dell'Osservatorio~2, I-35122 Padova, Italy\\
}

\date{Received 31 July 2002; accepted 16 September 2002}

\maketitle   
   
\begin{abstract}   
  
  We present surface photometry and stellar kinematics of a sample of
  5 SB0 galaxies: ESO 139-G009, IC 874, NGC 1308, NGC 1440 and NGC
  3412. We measured their bar pattern speed using the
  Tremaine-Weinberg method, and derived the ratio, $\vpd$, of the
  corotation radius to the length of the bar semi-major axis.  For all
  the galaxies, $\vpd$ is consistent with being in the range from 1.0
  and 1.4, \ie\ that they host fast bars.  This represents the largest
  sample of galaxies for which $\vpd$ has been measured this way.
  Taking into account the measured distribution of $\vpd$ and our
  measurement uncertainties, we argue that this is probably the true
  distribution of $\vpd$.  If this is the case, then the
  Tremaine-Weinberg method finds a distribution of $\vpd$ which is in
  agreement with that obtained by hydrodynamical simulations.
  We compared this result with recent high-resolution $N$-body
  simulations of bars in cosmologically-motivated dark matter halos,
  and conclude that these bars are not located inside centrally
  concentrated dark matter halos.

\end{abstract}   
   
\begin{keywords} 
  galaxies: elliptical and lenticular, cD --- 
  galaxies: haloes --- 
  galaxies: kinematics and dynamics --- 
  galaxies: photometry
\end{keywords}

\section{Introduction}   
\label{sec:introduction}   

The pattern speed of a bar, $\om$, is its main kinematic observable.
When parameterized by the distance-independent ratio $\vpd \equiv
\lag/\len$ (where $\lag$ is the Lagrangian/corotation radius, at which
a star is at rest in the bar's rest frame, and $\len$ is the bar
semi-major axis), it permits the classification of bars into fast
($1.0 \leq \vpd \leq 1.4$) and slow ($\vpd > 1.4$) ones. If $\vpd
  < 1.0$ orbits are elongated perpendicular to the bar, so that
  self-consistent bars cannot exist in this regime (Contopoulos 1980).
  
A robust method for measuring $\vpd$ relies on hydrodynamical
simulations to model gas, particularly at shocks.  These studies
find fast bars (\eg\ Lindblad \etal\ 1996; Lindblad \& Kristen 1996;
Weiner \etal\ 2001).  Hydrodynamical simulations can also obtain
$\vpd$ by matching morphological features in \hi\ (\eg\ Laine 1996;
England \etal\ 1990; Hunter \etal\ 1989; Aguerri \etal\ 2001).
Moreover, if the leading, offset dust lanes frequently found in bars
can be identified with shocks, then fast bars seem to be the norm in
late-type galaxies (van Albada \& Sanders 1982; Athanassoula 1992).
A model-independent method for measuring $\om$ directly was obtained
by Tremaine \& Weinberg (1984).  The Tremaine-Weinberg method
(hereafter, the TW method) is given by the simple expression $\pin
\om\sin i = \kin$, where $\pin$ and $\kin$ are luminosity-weighted
mean position and velocity measured along slits parallel to the
line-of-nodes.  If a number of slits at different offsets from the
major-axis are obtained for a galaxy, then plotting $\kin$ versus
$\pin$ for the different slits produces a straight line with slope
$\om \sin i$.  To date this method has been applied successfully to 3
early-type barred galaxies (Kent 1987; Merrifield \& Kuijken 1995;
Gerssen \etal\ 1999; Debattista \etal\ 2002) and preliminary results
have been obtained for another 5 galaxies ranging from SB0 to SBb
(Debattista \& Williams 2001; Gerssen 2002).  Although all these
galaxies are consistent with having fast bars, the sample size is
still small enough that the range of $\vpd$ recovered by the TW method
is still poorly defined.  We have started a program to enlarge the
sample of pattern speeds measured with the TW method, including to
galaxies of various bar strength, environment, luminosity, inclination
etc.  In a previous paper, (Debattista \etal\ 2002, hereafter Paper
I), we reported on the special case of NGC 1023, a system which shows
evidence of a past interaction with one or more of its satellite
galaxies.  In this paper, we present results for an additional 5 SB0
galaxies.

\section{The Sample}
\label{sec:sample}

Our preliminary sample of galaxies consisted of 11 SB0 objects
selected to be bright and undisturbed, with no evidence of dust or
spiral arms to complicate the TW analysis, generally not too many
nearby stars, and a bar at intermediate angle between the orientation
of the major and minor axes.
After detailed analysis of their photometrical properties (as
described in Section \ref{sec:photometry} below), 5 of them were
rejected due to either strong variation in the ellipticity and/or
position angle of the isophotal profiles at large galactocentric
distances, or due to the presence of dust in their inner regions. Of
the remaining 6, Paper I studied one of them, NGC 1023; in this paper
we will analyze the remaining 5 galaxies: ESO 139-G009 (SAB0), IC 874
(SB0), NGC 1308 (SB0), NGC 1440 (SB0), and NGC 3412 (SB0).
A compilation from the literature of their properties is given in
Table \ref{tab:properties}.

\begin{table*}
\caption[]{Parameters of the sample galaxies}  
\begin{center}  
\begin{tabular}{llcrcrccc}  
\hline  
\multicolumn{1}{c}{Galaxy} &  
\multicolumn{1}{c}{Type} &  
\multicolumn{1}{c}{$i$} &    
\multicolumn{1}{c}{PA} &    
\multicolumn{1}{c}{$B_T$} &  
\multicolumn{1}{c}{$D_{25} \times d_{25}$} &  
\multicolumn{1}{c}{$V_{{\rm CBR}}$} &  
\multicolumn{1}{c}{$D$} &  
\multicolumn{1}{c}{$M_{B_T}^0$} \\ 
\multicolumn{1}{c}{} &  
\multicolumn{1}{c}{(RC3)} &
\multicolumn{1}{c}{($^\circ$)} &    
\multicolumn{1}{c}{($^\circ$)} &    
\multicolumn{1}{c}{(mag)} &
\multicolumn{1}{c}{(arcsec)} &  
\multicolumn{1}{c}{(\kms)} &  
\multicolumn{1}{c}{(Mpc)} &  
\multicolumn{1}{c}{(mag)} \\  
\multicolumn{1}{c}{(1)} &  
\multicolumn{1}{c}{(2)} &  
\multicolumn{1}{c}{(3)} &  
\multicolumn{1}{c}{(4)} &  
\multicolumn{1}{c}{(5)} &  
\multicolumn{1}{c}{(6)} &  
\multicolumn{1}{c}{(7)} &  
\multicolumn{1}{c}{(8)} &  
\multicolumn{1}{c}{(9)} \\  
\hline  
ESO 139-G009  & (R)SAB0(rs) & $43.6_{-6.2}^{+5.1}$ &  
  94 & 14.35 & $ 74\times 60$ & 5389 & 71.9 & $-20.28$ \\
IC 874        & SB0(rs)     & $35.6_{-6.2}^{+5.0}$ &  
  17 & 13.60 & $ 60\times 43$ & 2602 & 34.7 & $-19.41$ \\
NGC 1308      & SB0(r)      & $43.6_{-8.0}^{+6.2}$ & 
 135 & 14.70 & $ 70\times 51$ & 6180 & 82.4 & $-19.88$ \\
NGC 1440      & (R')SB0(rs) & $40.7_{-5.0}^{+4.3}$ & 
  28 & 12.90 & $128\times 97$ & 1382 & 18.4 & $-18.90$ \\
NGC 3412      & SB0(s)      & $55.8_{-2.8}^{+2.6}$ & 
 155 & 11.43 & $218\times123$ & 1202 & 16.0 & $-19.73$ \\
\hline  
\noalign{\smallskip}  
\noalign{\smallskip}  
\noalign{\smallskip}  
\end{tabular}  
\begin{minipage}{18cm}  
NOTE -- 
Col.(2): Morphological classification from RC3.
Col.(3): Galaxy inclination from $\log{R_{25}}$ in RC3.
Col.(4): Major-axis position angle from RC3, except for 
         NGC 1308 and NGC 3412 (LEDA).  
Col.(5): Total observed blue magnitude from LEDA.
Col.(6): Apparent isophotal diameters measured at a surface-brightness 
         level of $\mu_B=25$ \mas\ from RC3.
Col.(7): Radial velocity with respect to the CMB radiation from RC3, 
         except for NGC 1308 (LEDA).
Col.(8): distance obtained as $V_{{\rm CBR}}/H_0$ with $H_0=75$ \kms\ 
         Mpc$^{-1}$.
Col.(9): Absolute total blue magnitude from $M_B$ 
         corrected for inclination and extinction as in LEDA
         and adopting $D$.
\end{minipage}  
\end{center}  
\label{tab:properties}  
\end{table*}

Four of these galaxies, ESO 139-G009, IC 874, NGC 1308, and NGC 1440
were previously poorly studied. Indeed no surface photometry or
stellar kinematics was available for the first three of these, and
only $V-$band photometry was available for NGC 1440 (Kodaira \etal\ 
1990), from which Baggett \etal\ (1998) obtained a decomposition into
bulge and disc.

This was not the case of NGC 3412, for which both surface photometry
and stellar kinematics were available.  Optical and NIR imaging of NGC
3412 were obtained by Kent (1984, $r$), Kodaira \etal\ (1990, $V$),
Shaw \etal\ (1995, {\it JHK\/}) and Ann (2001, {\it UBVRI\/} and
H$\alpha$).  Both Shaw \etal\ (1995) and Ann (2001) measured an
isophotal twist in the nuclear region of this galaxy, suggesting that
its bulge is triaxial.  Bulge-disc decompositions were perfomed by
Kent (1985, $r$) and Baggett \etal\ (1998, $V$).
Kuijken \etal\ (1996) obtained the position-velocity diagram of the
stellar component along $\rm PA = 155\degrees$ (the ``major-axis'',
but see our discussion in Section \ref{sec:photometry} below) of NGC
3412, to search for counter-rotation.  They found that the fraction of
counter-rotating disc stars is less than 5 per cent. Spatially
resolved stellar kinematics of this galaxy have been measured by
Fisher (1997, $\rm PA = 55^\circ,155^\circ$) and Neistein \etal\ 
(1999, $\rm PA = 155^\circ$), who derived the Tully-Fisher relation 
for a sample of nearby lenticular galaxies.

\section{Surface Photometry}
\label{sec:photometry}

The photometric observations of ESO 139-G009 and IC 874 were carried
out with the 3.5-m New Technology Telescope (NTT) at the European
Southern Observatory (ESO) in La Silla (Chile) in May 2001. We used
the ESO Multi-Mode Instrument (EMMI) in red imaging and low-dispersion
spectroscopic (RILD) mode.  The detector was the No. 36 Tektronik
TK2048 EB CCD with $2048 \times 2048$ pixels. The image scale was 0.27
arcsec pixel$^{-1}$.
We imaged NGC 1308 and NGC 3412 at the 1-m Jacobus Kapteyn Telescope
located at the Roque de los Muchachos Observatory (ORM) in La Palma
(Spain) in October 2001 and May 2000, respectively. We used a SITe2
CCD with $2048 \times 2048$ pixels and an image scale of 0.33 arcsec
pixel$^{-1}$.
Finally, we observed NGC 1440 with the Danish 1.54-m telescope at ESO
in La Silla in November 2001. The telescope mounted the Danish Faint
Object Spectrograph and Camera (DFOSC) with the EEV/MAT 44-82 CCD with
$2048 \times 4096$ pixels. The image scale was 0.39 \asecpix. The log
of the photometrical observations is given in Table
\ref{tab:login_phot}.

All images were reduced using standard \iraf\footnote{\iraf~is
  distributed by NOAO, which is operated by AURA Inc., under contract
  with the National Science Foundation} tasks.  First, we subtracted a
bias frame consisting of 10 exposures for each night.  The images were
then flat-fielded using sky flats taken at twilight.  The
sky-background level was removed by fitting a second order polynomial
to the regions free of sources in the images. Special care was taken
at this point to obtain a final background in the images free of
sources, in order to reach the outermost parts of the objects.  Cosmic
rays were removed by combining the different exposures for each filter
using a sigma clipping rejection algorithm.

\begin{figure*}
 \leavevmode{\psfig{figure=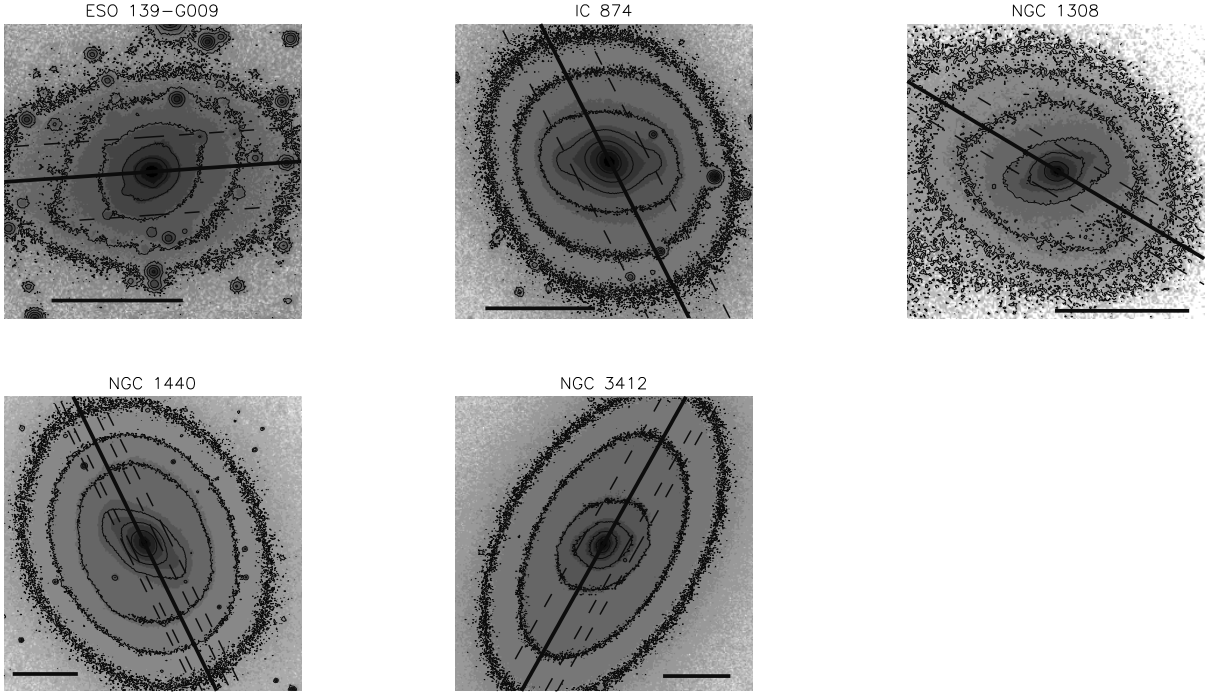,width=18.truecm}}
\caption[]{$I$-band images of the sample galaxies.  
  The isocontours are spaced at 1 mag arcsec$^{-2}$ and the outermost
  corresponds to $\mu_I = 23$ \mas.  In each panel, North is up and
  East is left, and the bottom horizontal line is 30 arcsec long. For
  each galaxy, the thick continuous line and the dashed lines show the
  slit position along major-axis and the offsets in the spectroscopic
  observations.}
\label{fig:images}
\end{figure*}

Photometric calibrations were achieved by means of standard stars.
The calibration constants include corrections for atmospheric and
Galactic extinction, and a colour term.  Fig.  \ref{fig:images} shows
calibrated $I$-band images of the galaxies.

\begin{table}   
\caption{Log of the photometrical observations}
\begin{flushleft}   
\begin{tabular}{l r r r r}   
\hline   
\multicolumn{1}{c}{Galaxy} &   
\multicolumn{1}{c}{Date} &  
\multicolumn{3}{c}{Exp. Time (sec)} \\ 
\noalign{\smallskip}  
\cline{3-5}
\noalign{\smallskip}  
\multicolumn{1}{c}{} &  
\multicolumn{1}{c}{} &  
\multicolumn{1}{c}{$B$} &  
\multicolumn{1}{c}{$V$} &  
\multicolumn{1}{c}{$I$} \\
\hline  
ESO 139-G009& 22-23 May 2001 & $1440^a$ &  $480^a$ &  $300^b$\\ 
IC  874     & 22-23 May 2001 &  $600^a$ &          &  $300^b$\\
NGC 1308    &    11 Oct 2001 &          &          & $7200^c$\\ 
NGC 1440    &    11 Nov 2001 & $2400^a$ &          & $3600^b$\\ 
NGC 3412    & 27-28 May 2000 & $2400^c$ & $2700^c$ & $3600^c$\\ 
\hline
\noalign{\smallskip}  
\end{tabular}  
\begin{minipage}{8.5cm}  
NOTE -- $^a$ Bessel, $^b$ Gunn, $^c$ Harris
\end{minipage}  
\end{flushleft}  
\label{tab:login_phot}   
\end{table}

\section{BAR LENGTH}

\subsection{Photometrical parameters}
\label{sec:photometrical_parameters}

\begin{figure*}
  \leavevmode{\psfig{figure=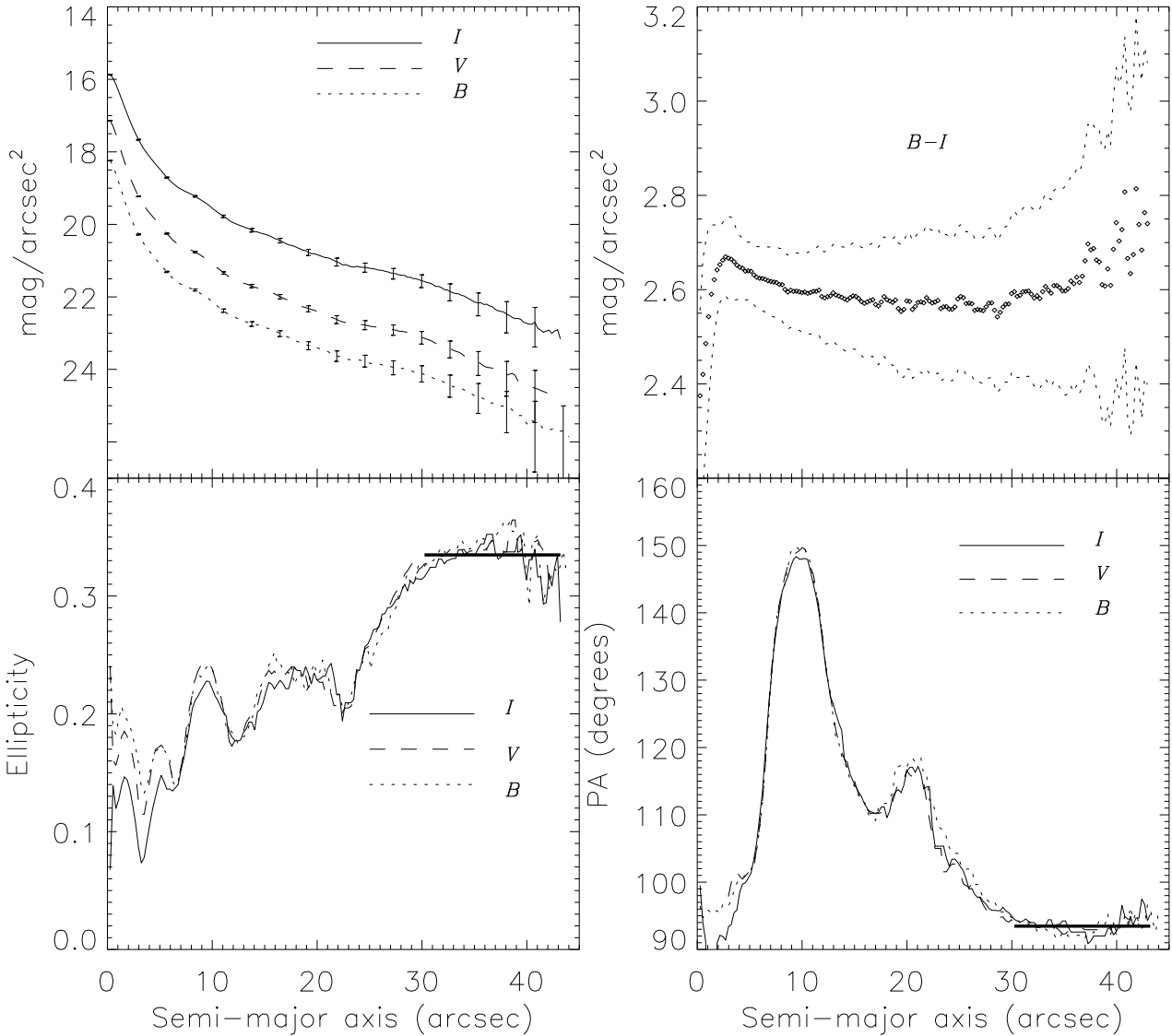,height=17.truecm,width=17.truecm}}
\caption[]{Surface brightness,  ellipticity, position angle and $B-I$
  colour radial profiles of ESO 139-G009. Continuous, dashed and
  dotted line refer to $I$, $V$ and $B-$band data, respectively.  The
  thick lines represent the fits to the $I$-band ellipticity and
  position angle of the galaxy in the disc region.}
\label{fig:eso139}
\end{figure*}

\begin{figure*}
 \leavevmode{\psfig{figure=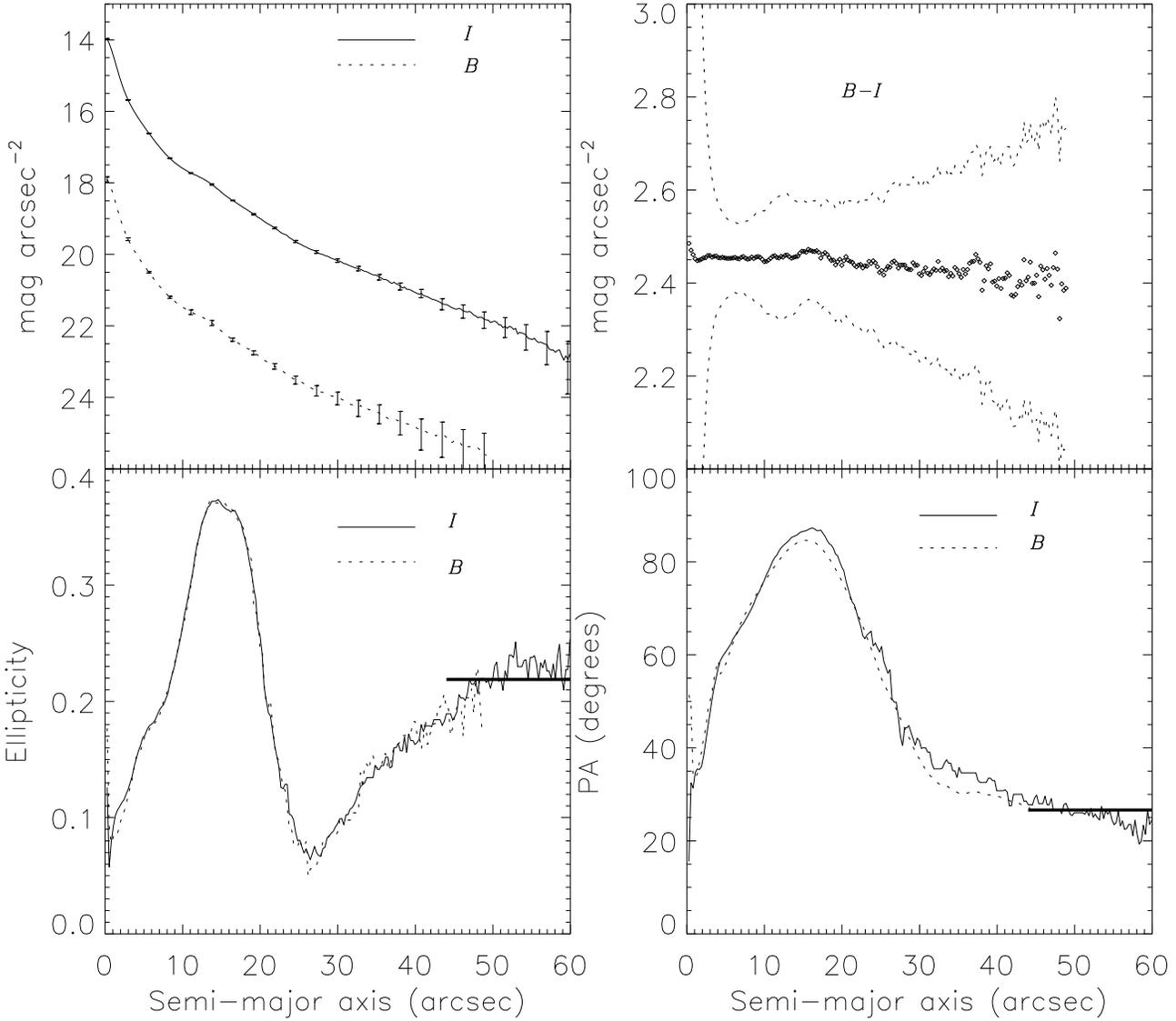,height=17.truecm,width=17.truecm}}
\caption[]{As in Fig. \ref{fig:eso139} but for IC 874.} 
\label{fig:ic874}
\end{figure*}

\begin{figure*}
 \leavevmode{\psfig{figure=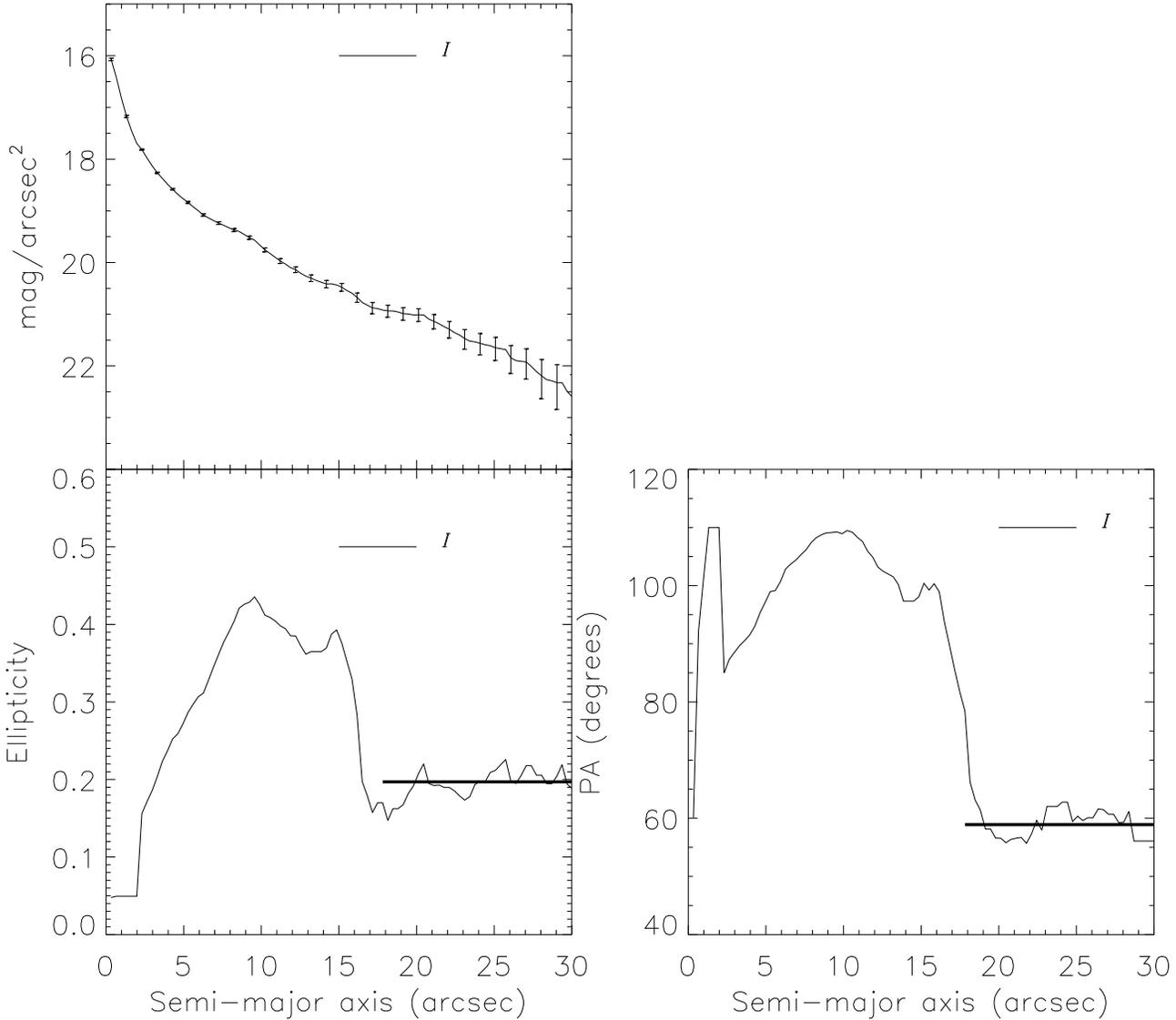,height=17.truecm,width=17.truecm}}
\caption[]{As in Fig. \ref{fig:eso139} but for NGC 1308. Only
  $I-$band data are available for this galaxy and therefore no $B-I$
  colour radial profile is provided.}
\label{fig:n1308}
\end{figure*}

\begin{figure*}
 \leavevmode{\psfig{figure=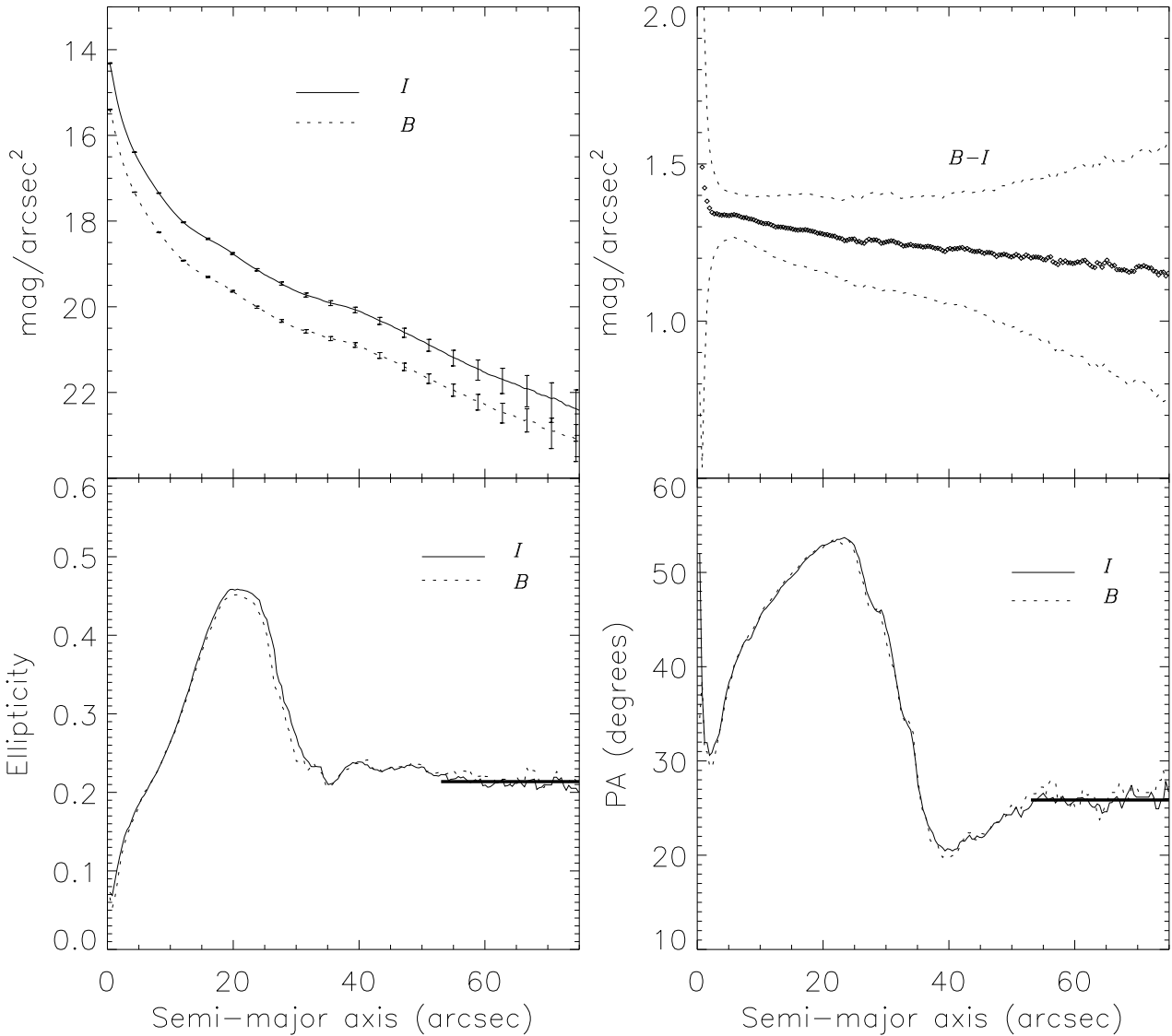,height=17.truecm,width=17.truecm}}
\caption[]{As in Fig. \ref{fig:eso139} but for NGC 1440.} 
\label{fig:n1440}
\end{figure*}

\begin{figure*}
 \leavevmode{\psfig{figure=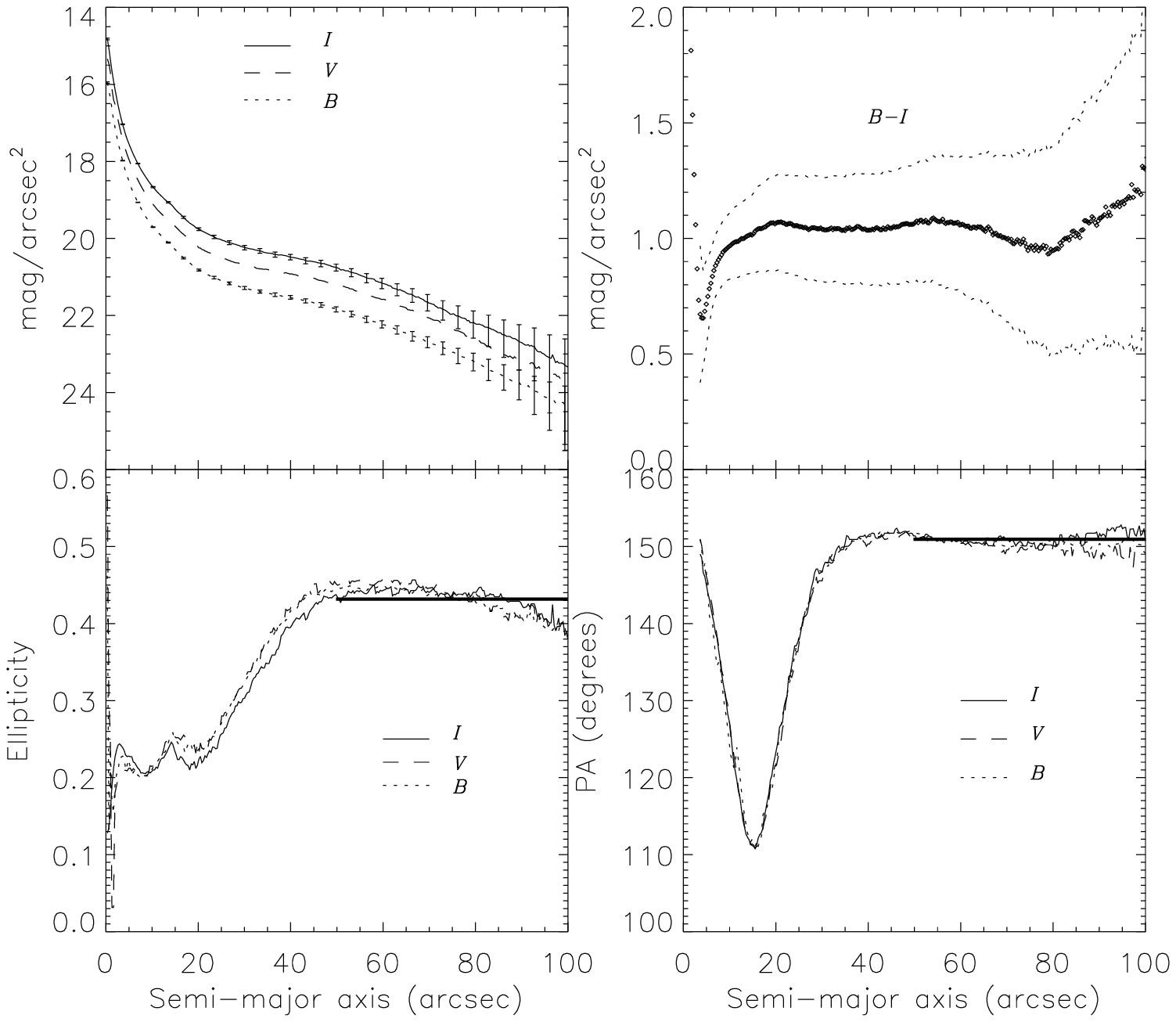,height=17.truecm,width=17.truecm}}
\caption[]{As in Fig. \ref{fig:eso139} but for NGC 3412.} 
\label{fig:n3412}
\end{figure*}

Isophote-fitting with ellipses, after masking foreground stars and bad
pixels, was carried out using the \iraf\ task {\tt ELLIPSE}.  In all
cases, we first fit ellipses allowing their centers to vary; within
the errors, no variation in the ellipse centers was found for the 5
galaxies studied in this paper. Since patchy obscuration can cause
variations of ellipse centers, we conclude that there is no evidence
for such obscuration in our sample, as we also verified from colour
maps and $B-I$ colour profiles, where possible.  The final ellipse
fits were done at fixed ellipse centers.  In Figs.
\ref{fig:eso139}-\ref{fig:n3412}, we show the resulting surface
brightness, $B-I$ colour, ellipticity and position angle profiles of
the galaxies.  The inclination and position angle of the galaxies were
determined by averaging the outer isophotes, as shown in Fig.
\ref{fig:images}.  The values obtained are reported in Table
\ref{tab:measured_pa}. The inclinations obtained for the galaxies are,
within the errors, consistent with the values reported in the
literature.  The position angle has larger differences, the biggest
being $\Delta {\rm PA} = 99\degrees$ in the case of NGC 1308.  We
speculate that the LEDA value of the position angle for this galaxy is
contaminated by the bright star to its east.  In measuring the
photometrical parameters of this galaxy, we masked out the whole
eastern part, to minimize the contamination from this star.
In the case of NGC 3412, it is possible to compare our values of $i$
and $\rm PA$ with the ones obtained by Kent (1984), who found
$i=56\fdg6$ and $\rm PA=151\fdg6$.  These data are in very good
agreement with our measurements, as given in Tab.
\ref{tab:measured_pa}.

Using the ellipse fits, we measured the angle, $\psi_{\rm bar}$,
between the bar major-axis and the line-of-nodes in the deprojected
plane of the disc.  The values of this parameter are reported in Tab.
\ref{tab:measured_pa}.

\begin{table}   
\caption{Inclination, disc position angle, and intrinsic bar angle
relative to the line-of-nodes of the sample galaxies}
\begin{center}   
\begin{tabular}{l c c c}   
\hline   
\multicolumn{1}{c}{Galaxy} &   
\multicolumn{1}{c}{$i$} &
\multicolumn{1}{c}{P.A.$_{{\rm disc}}$} &
\multicolumn{1}{c}{$\psi_{\rm bar}$} \\
\multicolumn{1}{c}{} &     
\multicolumn{1}{c}{($\degrees$)} &
\multicolumn{1}{c}{($\degrees$)} &
\multicolumn{1}{c}{($\degrees$)} \\ 
\hline 
ESO 139-G009 & $46.2\pm1.5$ & $ 93.5\pm1.0$ & $77\pm1$ \\ 
IC 874       & $38.7\pm1.8$ & $ 26.6\pm2.7$ & $70\pm3$ \\ 
NGC 1308     & $35.9\pm2.6$ & $ 58.9\pm4.1$ & $60\pm1$ \\
NGC 1440     & $37.8\pm0.9$ & $ 25.9\pm0.6$ & $40\pm1$ \\ 
NGC 3412     & $55.3\pm2.1$ & $151.0\pm0.9$ & $84\pm2$ \\ 
\hline
\end{tabular}  
\label{tab:measured_pa}   
\end{center}   
\end{table}

\subsection{Bar semi-major axis length}

Determining the length of a bar is not entirely trivial.  This is
particularly true in early-type galaxies, for which there is usually
no obvious spiral structure or star formation beyond the bar marking
its end, so that the luminosity distribution of the bar gradually
melds into that of the disc.  The presence of a large bulge may
complicate further the measurement of $\len$.
Fourier decomposition of the galaxy light (Ohta et al 1990; Aguerri et
al 2000), identification of a change in the slope of the
surface-brightness profile along the bar major-axis (Chapelon et al.
1992), visual ispection of galaxy images (Martin 1995), location of
minimum/maximum in the ellipticity profile (Wozniak et al. 1995), and
modeling the structural components contributing to the galaxy surface
brightness (Prieto et al 2001) are among the several methods which
have been proposed for measuring $\len$.
We have used three of these, yielding 4 estimates of $\len$ ($a_{B,1}$
to $a_{B,4}$).  The mean of these is our best estimate of $\len$ and
we use the largest deviations from the mean for the error estimates.
The values obtained are given in Table \ref{tab:bars_length}.  The
three methods are the following:

\begin{enumerate}
\item{\it Fourier amplitudes}. Since a bar represents a bisymmetric
  departure from axisymmetry, a Fourier decomposition is a natural way
  to measure $\len$.  The method of Aguerri \etal\ (2000) is based on
  the ratios of the intensities in the bar and the inter-bar region.
  The azimuthal surface brightness profile of the deprojected galaxies
  is decomposed in a Fourier series.  The bar intensity, $I_{b}$, then
  is defined as: $I_{b}=I_{0}+I_{2}+I_{4}+I_{6}$ (where $I_{0},
  I_{2},I_{4}$ and $I_{6}$ are the $m=0$, 2, 4 and 6 terms of the
  Fourier decomposition, respectively).  The inter-bar intensity is
  defined as: $I_{ib}=I_{0}-I_{2}+I_{4}-I_{6}$.  The bar region is
  defined as the region where $I_{b}/I_{ib}>0.5[\max(I_{b}/I_{ib})-
  \min(I_{b}/I_{ib})]+\min(I_{b}/I_{ib})$; thus the semi-major axis of
  the bar is identified as the outer radius at which
  $I_{b}/I_{ib}=0.5[\max(I_{b}/I_{ib})-\min(I_{b}/I_{ib})]+\min(I_{b}/I_{ib})$.
  Athanassoula \& Misiriotis (2002) applied this method to the
  analytic models of Athanassoula (1992) (for which $\len$ are known
  exactly) and found an accuracy of better than 8 per cent.  In
  practical applications to early-type galaxies, adjustment for the
  presence of a massive bulge, which in deprojection may result in a
  higher peak in $I_{b}/I_{ib}$ than that due to the bar, may be
  required (\eg\ Paper I).  The values listed as $a_{B,1}$ in Table
  \ref{tab:bars_length} are the ones which were obtained using this
  method. Figure \ref{fig:profiles} shows relative amplitudes of the
  $m=2,4,6$ Fourier components and the bar/interbar intensity ratio of
  the galaxies.
  
\item{\it Fourier and ellipse phases}. In their analysis of $N$-body
  models, Debattista \& Sellwood (2000) also made use of a Fourier
  decomposition of the surface density.  Here we adopt their $m=2$
  (deprojected) phase method for $a_{B,2}$, namely that a bar can
  extend only as far out as the phase of the $m=2$ moment is constant
  (within the errors, which here must include uncertainties in PA and
  $i$).  In $N$-body simulations, secondary structures, such as rings
  or spirals, sometimes lead to an over-estimate of the semi-major
  axis.  However, our early-type galaxies were selected to be free of
  spirals and rings.  A related, but not equivalent, method uses the
  phases of the ellipse fits, which then gives us estimate $a_{B,3}$
  in Table \ref{tab:bars_length}. In Figure \ref{fig:profiles}, we
  plot the phases of the Fourier $m=2$ moments and of the ellipses,
  showing $a_{B,2}$ and $a_{B,3}$.
  
\item{\it Decomposition of surface brightness profiles}. The surface
  brightness profiles of galaxies can be decomposed into a number of
  structural components, each of which may be described by some simple
  analytic model.  Barred galaxy models are complicated by their
  non-axisymmetric nature; a method for such decompositions has been
  developed by Prieto \etal\ (2001), which we used here.  We assumed
  four different components: bulge, disc, bar and lens. The bulge
  luminosity was modeled with a S\'{e}rsic's law (S\'{e}rsic 1968), while the
  disc was modelled by an exponential profile (Freeman 1970).  Two
  profile types for the bars were used: elliptical (Freeman 1966) and
  flat (Prieto \etal\ 1997). The best fit for NGC 3412 was obtained
  with an elliptical bar; for the other galaxies, we used flat bars.
  The lenses were characterized by a smooth luminosity gradient with a
  very sharp cut-off (Duval \& Athanassoula 1983).  The model of NGC
  1308 did not need to include a lens; for all the other galaxies, the
  fit was improved by including one.
  
  In all, ten free parameters needed to be fit: the central surface
  brightness and the scale-length of the disc, the effective surface
  brightness, the effective radius, and the S\'{e}rsic index of the bulge,
  the central surface brightness, the semi-major and the semi-minor
  axes of the bar, the central surface brightness and the scale-radius
  of the lens.  The fits were performed on the major and minor-axes of
  the bar, after deprojecting the image by means of a flux-conserving
  stretch along the disc minor-axis.
  
  We used an interactive, iterative method, described in Prieto \etal\ 
  (2001) to obtain initial estimates of all the parameters.  Then,
  these estimates of the parameters were used as input parameters in
  an automatic fitting routine using a Levenberg-Marquardt nonlinear
  fitting algorithm (Press \etal\ 1992).  Figure \ref{fig:profiles}
  shows the best fit of the surface-brightness profiles along the
  major axis of the bar for the galaxies of our sample.  Table
  \ref{tab:bars_length} also includes the bar semi-major axis 
  measured with this method ($a_{B,4}$).
\end{enumerate}

\begin{table*}   
\begin{center}   
\caption{Bar semi-major axis of the galaxies}
\begin{tabular}{l c c c c c}   
\hline  
\multicolumn{1}{c}{Galaxy} & 
\multicolumn{1}{c}{$a_{B,1}$} &
\multicolumn{1}{c}{$a_{B,2}$} &   
\multicolumn{1}{c}{$a_{B,3}$} &
\multicolumn{1}{c}{$a_{B,4}$} &  
\multicolumn{1}{c}{$<a_{B}>$} \\
\multicolumn{1}{c}{} & 
\multicolumn{1}{c}{(arcsec)} &
\multicolumn{1}{c}{(arcsec)} &   
\multicolumn{1}{c}{(arcsec)} &
\multicolumn{1}{c}{(arcsec)} &  
\multicolumn{1}{c}{(arcsec)} \\
\hline 
ESO 139-G009 & 23.4 & 14.0 & 16.1 & 14.4 & $17.0^{+6.4}_{-3.0}$\\ 
IC 874       & 18.7 & 21.0 & 25.0 & 15.0 & $19.9^{+5.1}_{-4.9}$\\ 
NGC 1308     & 13.2 &  9.0 & 13.2 & 14.2 & $12.4^{+1.8}_{-3.4}$\\ 
NGC 1440     & 23.2 & 24.6 & 30.5 & 19.2 & $24.4^{+6.1}_{-5.2}$\\  
NGC 3412     & 28.2 & 30.0 & 32.2 & 34.0 & $31.1^{+2.9}_{-2.9}$\\ 
\hline
\label{tab:bars_length}   
\end{tabular}   
\end{center}   
\end{table*}

\section{Long-Slit Spectroscopy}
\label{sec:spectroscopy}

The spectroscopic observations of ESO 139-G009, IC 874, NGC 1308 and
NGC 1440 were carried out with NTT at ESO in La Silla in May and
November 2001.  The NTT mounted EMMI in red medium-dispersion
spectroscopic (REMD) mode, using the grating No.~6 with 1200 $\rm
grooves\,mm^{-1}$ in first order with a $1.0$ arcsec $\times$ $5.5$
arcmin slit. The detector was the No. 36 Tektronix TK2048 EB CCD with
$2048\,\times\,2048$ pixels of $24\,\times\,24$ $\rm \mu m^2$.  It
yielded a wavelength coverage between about 4840 \AA\ and 5490 \AA\ 
with a reciprocal dispersion of 0.320 $\rm \AA\,pixel^{-1}$.  The
instrumental resolution was $1.19$ \AA\ (FWHM) corresponding to
$\sigma_{\it inst}\approx30$ \kms\ at 5170 \AA . The spatial scale was
$0.270$ arcsec pixel$^{-1}$.

We observed NGC 3412 with the 3.6-m Telescopio Nazionale Galileo (TNG)
at the ORM in La Palma in February 2001. The TNG was equipped with the
Low Resolution Spectrograph (DOLORES); we used the HR-V grism No.~6
with 600 $\rm grooves\,mm^{-1}$ in combination with the $0.7$ arcsec
$\times$ $8.1$ arcmin slit and the Loral CCD with $2048\,\times\,2048$
pixels of $15\,\times\,15$ $\rm \mu m^2$.  The wavelength range
between about 4700 and 6840 \AA\ was covered with a reciprocal
dispersion of 1.055 \AA\ pixel$^{-1}$.  The instrumental resolution
(obtained by measuring the width of emission lines of a comparison
spectrum after the wavelength calibration) was $3.10$ \AA\ (FWHM).
This corresponds to an instrumental dispersion of $\sigma_{\it
  inst}\approx80$ \kms\ at 5170 \AA.  The spatial scale along the slit
was $0.275$ arcsec pixel$^{-1}$.

\begin{figure*}
 \leavevmode{\psfig{figure=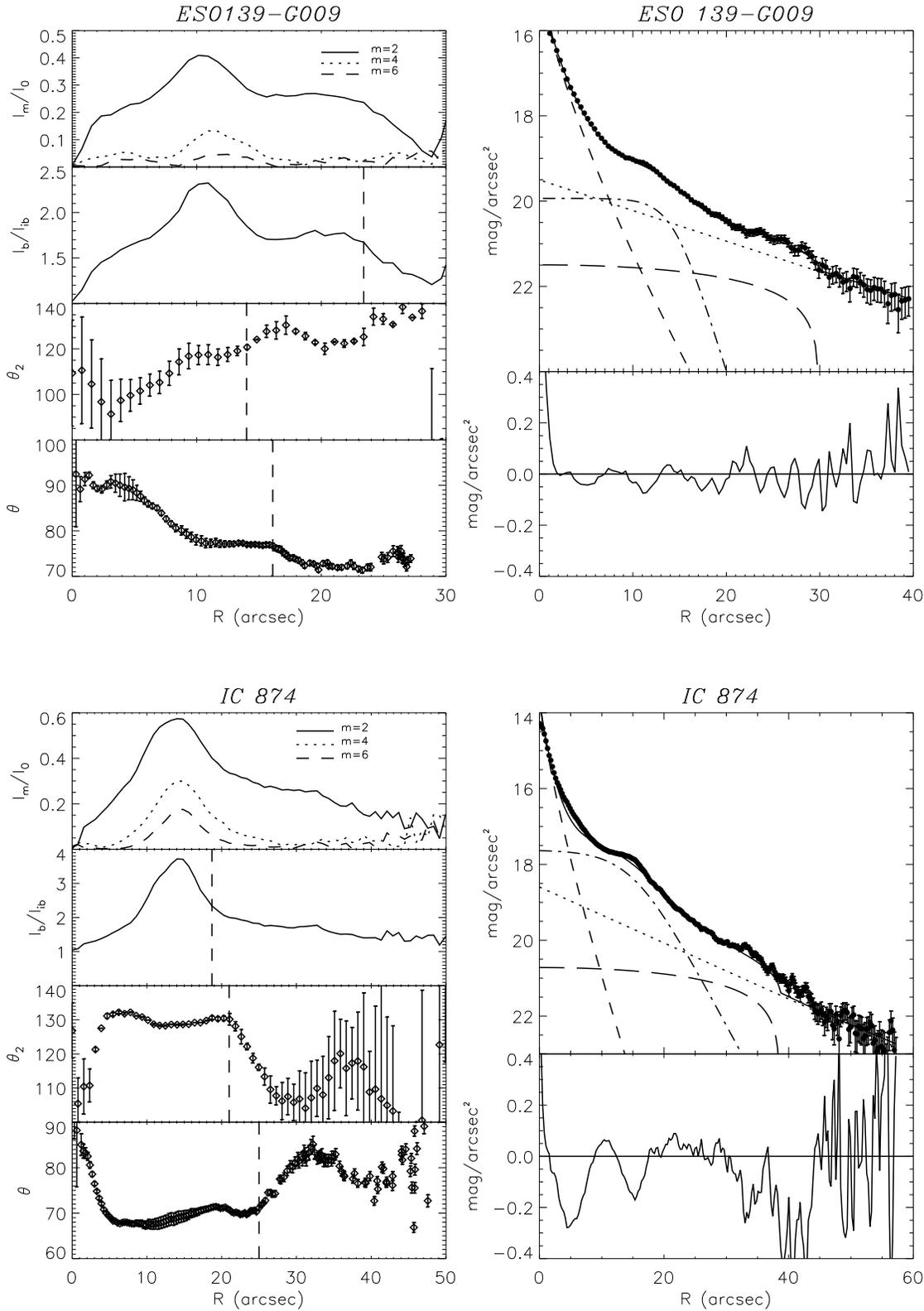,width=15.truecm}}
\caption[]{Bar semi-major axis length of the sample galaxies.
  Left panels (from top to bottom): Relative amplitudes of the
  $m=2,4,6$ Fourier components, bar/interbar intensity ratio, phase
  angle of the $m=2$ Fourier component, and position angle of the
  deprojected isophotal ellipses. The dashed vertical lines show the
  bar length ($a_{B,1}$, $a_{B,2}$, and $a_{B,3}$ of Table 4).
  Right panels (from top to bottom): decomposition of the radial
  surface brightness profiles in $I$-band along the bar major-axis,
  and residuals of the fit. The adopted structural components are
  bulge (dashed line), disc (dotted line), bar (dash-dotted line) and
  lens (long-dashed line). The continuous line represents the total
  model.}
\label{fig:profiles}
\end{figure*}

\addtocounter{figure}{-1}
\begin{figure*}
 \leavevmode{\psfig{figure=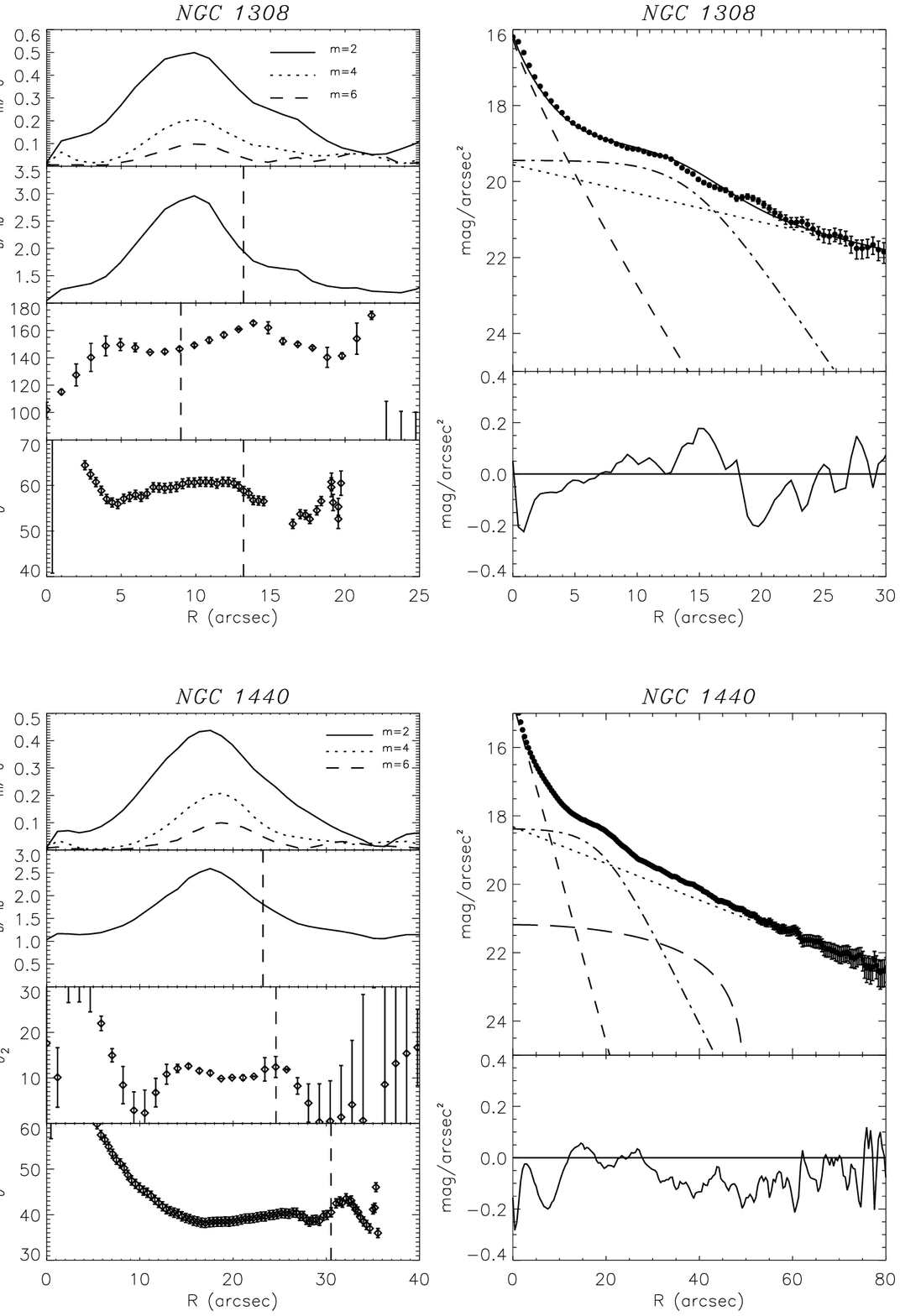,width=15.truecm}}
\caption[]{(continued)}
\label{fig:profiles}
\end{figure*}

\addtocounter{figure}{-1}
\begin{figure*}
 \leavevmode{\psfig{figure=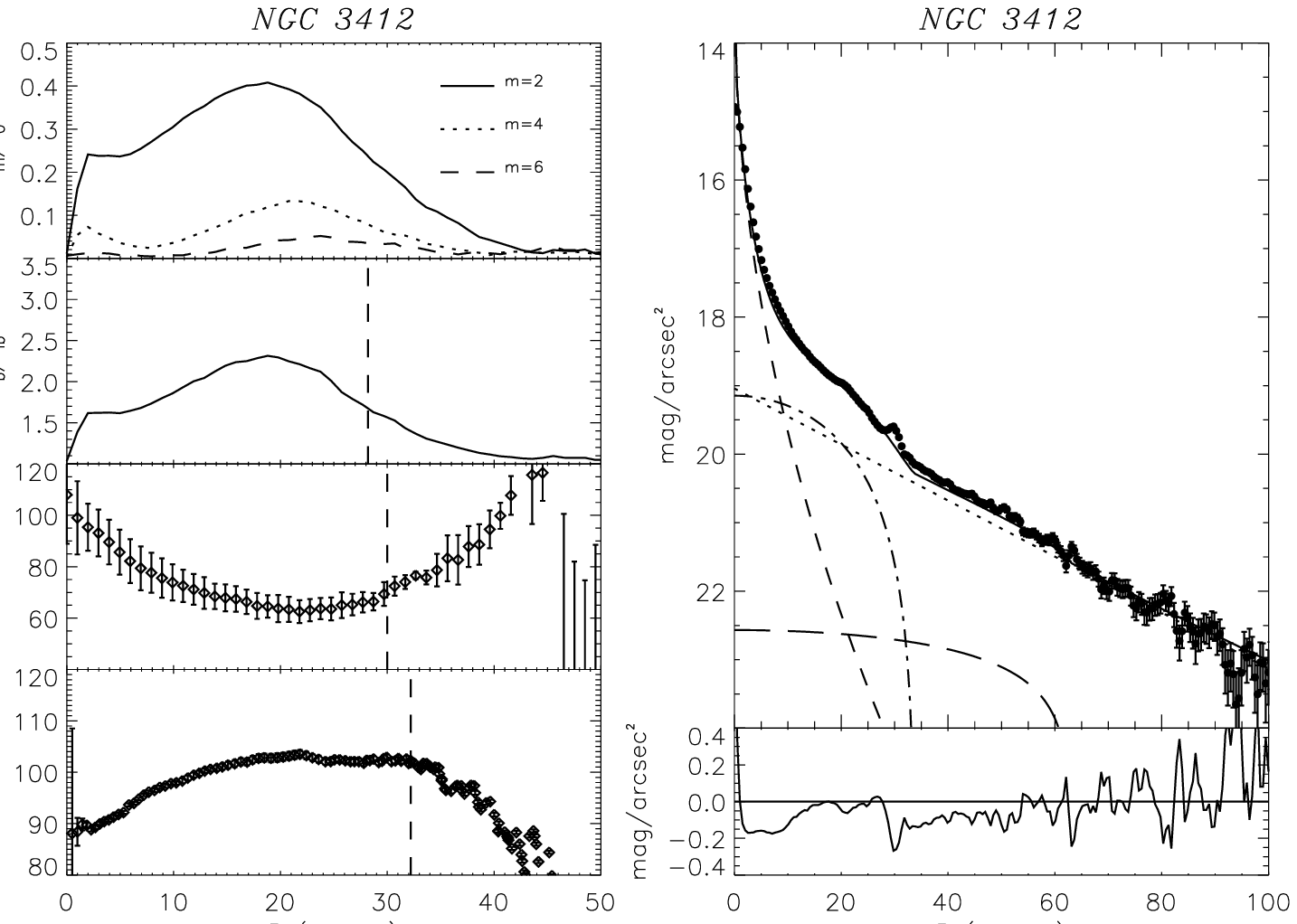,width=15.truecm}}
\caption[]{(continued)}
\label{fig:profiles}
\end{figure*}

\begin{table}   
\caption{Log of the spectroscopic observations}
\begin{center}   
\begin{tabular}{l r l c}   
\hline   \multicolumn{1}{c}{Galaxy}   &   
\multicolumn{1}{c}{Date}   & \multicolumn{1}{c}{Slit}   &    
\multicolumn{1}{c}{Exp.   Time}    \\
\multicolumn{1}{c}{}  & \multicolumn{1}{c}{} & \multicolumn{1}{c}{} &
\multicolumn{1}{c}{(sec)} \\ 
\hline 
ESO 139-G009  & 24 May 2001 & $10$ arcsec N & $7200$ \\ 
              & 23 May 2001 & major axis    & $5400$ \\ 
              & 23 May 2001 & $ 8$ arcsec S & $7200$ \\  
\hline 
IC 874        & 23 May 2001 & $10$ arcsec E & $10800$ \\ 
              & 22 May 2001 & major axis    & $5400$ \\  
              & 24 May 2001 & $10$ arcsec W & $10800$ \\ 
\hline 
NGC 1308      & 15 Nov 2001 & $ 5$ arcsec E & $10800$ \\ 
              & 17 Nov 2001 & major axis    & $5400$ \\ 
              & 16 Nov 2001 & $ 5$ arcsec W & $10800$ \\ 
\hline 
NGC 1440      & 17 Nov 2001 & $ 9$ arcsec E & $7200$ \\ 
              & 15 Nov 2001 & $ 6$ arcsec E & $5400$ \\ 
              & 15 Nov 2001 & major axis    & $5400$ \\ 
              & 16 Nov 2001 & $ 6$ arcsec W & $5400$ \\ 
              & 17 Nov 2001 & $11$ arcsec W & $7200$ \\             
\hline 
NGC  3412     & 01 Feb 2001 & $10$ arcsec E & $7200$ \\  
              & 02 Feb 2001 & major axis    & $7200$ \\ 
              & 01 Feb 2001 & $10$ arcsec W & $7200$ \\ 
              & 02 Feb 2001 & $15$ arcsec W & $7200$ \\ 
\hline
\label{tab:log_spec}   
\end{tabular}   
\end{center}   
\end{table}

For each galaxy, we took a spectrum with the slit along the major
axis, as well as two or more offset spectra with the slit parallel to
the major axis on each side of the center.  The major axis position
angles were chosen according to the values obtained from the surface
photometry, as described in Sect.  \ref{sec:photometrical_parameters}
and listed in Table \ref{tab:measured_pa}.  The location of the offset
spectra was chosen, using the photometry, to give large $\pin$ (and
therefore large $\kin$) and high $S/N$.  Integration time of the
galaxy spectra was split into exposures of 2700/3600 sec to deal with
cosmic rays.  A log of the observations is given in Table
\ref{tab:log_spec}.  In each observing run we obtained spectra of
twelve giant stars with spectral type ranging from late-G to early-K,
selected from Faber et al.  (1985), to use as templates in measuring
stellar kinematics.  Arc lamp spectra were taken before and/or after
every object exposure to allow an accurate wavelength calibration.
The range of the seeing FWHM during the observing runs was
$0.7$--$1.6$ arcsec in February 2001 (as measured by fitting a
two-dimensional Gaussian to the TNG guide star), $0.5$--$2.2$ arcsec
in May 2001, and $0.5$--$1.5$ arcsec in November 2001 (as measured by
the ESO Differential Image Motion Monitor).
 
All the spectra were bias subtracted, flatfield corrected, cleaned of
cosmic rays, corrected for bad pixels and columns, and wavelength
calibrated using standard {\tt MIDAS}\footnote{{\tt MIDAS} is
  developed and maintained by the European Southern Observatory}
routines as in Paper I. We checked that the wavelength rebinning was
done properly by measuring the difference between the measured and
predicted wavelengths (Osterbrock et al.  1996) for the brightest
night-sky emission lines in the observed spectral ranges.  The
resulting accuracy in the wavelength calibration is better than 2
\kms.

The spectra taken along the same axis for the same galaxy were
co-added using the center of the stellar continuum as reference.  The
contribution of the sky was determined from the outermost $\sim30$
arcsec at the two edges of the resulting spectra, where the galaxy
light was negligible, and then subtracted, giving a sky subtraction
better than 1 per cent. A one-dimensional sky-subtracted spectrum was
obtained for each kinematical template star.
 
\begin{figure*}
\leavevmode{\psfig{figure=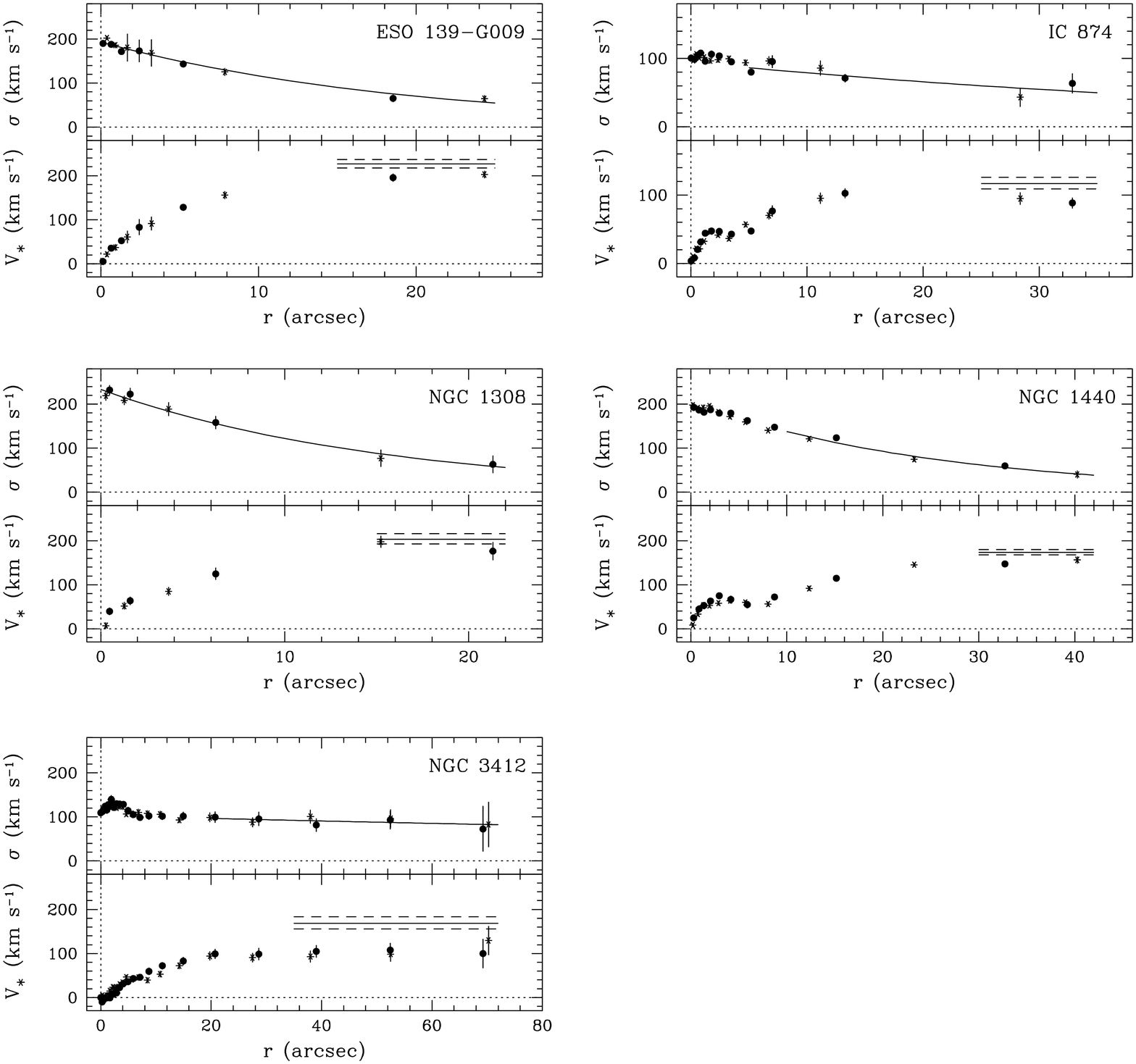,width=16.truecm}}
\caption[]{The   major-axis   radial   profiles  of   the   stellar
  line-of-sight velocity dispersion and velocity (after subtraction of
  the systemic velocity) of the sample galaxies. The profiles are
  folded around the center.  Filled dots and asterisks refer to
  receding and approaching side of the galaxy, respectively.  For each
  galaxy the exponential radial profile fitting the line-of-sight
  velocity dispersion profile and circular velocity obtained by
  applying asymmetric drift correction are plotted in upper and lower
  panels, respectively.  Note that all data at $|\phi| \leq
  30\degrees$ from all slits have been used for the fits to $V_{\rm
    c,flat}$, but only points on the major-axis are shown in this
  figure.}
\label{fig:asym_drift_corr}
\end{figure*}

\section{Bar and disc kinematics}
\label{sec:pattern_speeds}

\subsection{Stellar kinematics} 
\label{sec:kinematics}

We measured the stellar kinematics from the galaxy absorption features
present in the wavelength range centered on the Mg line triplet
($\lambda\lambda\,5164,5173,5184$ \AA, see Tab.  \ref{tab:TW_values})
using the Fourier Correlation Quotient method (Bender 1990; Bender et
al. 1994), as done in Paper I.
We adopted HR 6817 (K1III) as the kinematical template to measure the
stellar kinematics of NGC 3412, HR 7429 (K3III) for the kinematics of
ESO 139-G009 and IC 874, and HR 3145 (K2III) for the kinematics of NGC
1308 and NGC 1440.

The values of line-of-sight velocity $v$, and velocity dispersion
$\sigma$ measured along the different slits for each sample galaxy are
given in Table \ref{tab:kinematics}.

\begin{landscape}
\begin{table}   
\caption{Stellar kinematics of the sample galaxies.}   
\begin{tiny}
\begin{tabular}{rrr rrr rrr rrr}
\hline
\multicolumn{1}{c}{$r$} &   
\multicolumn{1}{c}{$v$} &     
\multicolumn{1}{c}{$\sigma$} &   
\multicolumn{1}{c}{$r$} &   
\multicolumn{1}{c}{$v$} &     
\multicolumn{1}{c}{$\sigma$} &  
\multicolumn{1}{c}{$r$} &   
\multicolumn{1}{c}{$v$} &     
\multicolumn{1}{c}{$\sigma$} &   
\multicolumn{1}{c}{$r$} &   
\multicolumn{1}{c}{$v$} &     
\multicolumn{1}{c}{$\sigma$} \\   
\multicolumn{1}{c}{(arcsec)} &   
\multicolumn{1}{c}{(km s$^{-1}$)} &      
\multicolumn{1}{c}{(km s$^{-1}$)} &
\multicolumn{1}{c}{(arcsec)} &   
\multicolumn{1}{c}{(km s$^{-1}$)} &      
\multicolumn{1}{c}{(km s$^{-1}$)} &
\multicolumn{1}{c}{(arcsec)} &   
\multicolumn{1}{c}{(km s$^{-1}$)} &      
\multicolumn{1}{c}{(km s$^{-1}$)} &
\multicolumn{1}{c}{(arcsec)} &   
\multicolumn{1}{c}{(km s$^{-1}$)} &      
\multicolumn{1}{c}{(km s$^{-1}$)} \\
\hline 
\multicolumn{3}{c}{ESO 139-G009 - 10 arcsec N}&   $  7.23$ & $2357.5\pm 4.2$ & $ 64.2\pm 4.1$  &   $ -0.22$ & $1592.4\pm 4.7$ & $197.6\pm 4.9$  &    $ -3.98$ & $ 815.6\pm 6.1$ & $123.2\pm 7.3$ \\ 
          &                 &                 &   $ 15.78$ & $2373.3\pm 7.8$ & $ 74.7\pm 7.0$  &   $  0.31$ & $1625.8\pm 4.4$ & $192.6\pm 4.1$  &    $ -3.43$ & $ 820.3\pm 5.9$ & $123.8\pm 6.6$ \\ 
 $-18.78$ & $5262.1\pm23.9$ & $ 90.3\pm14.1$  &   $ 35.47$ & $2395.4\pm13.0$ & $ 53.0\pm11.7$  &   $  0.85$ & $1646.0\pm 5.2$ & $186.6\pm 5.2$  &    $ -2.88$ & $ 826.9\pm 5.1$ & $120.6\pm 5.8$ \\      
 $ -3.68$ & $5321.0\pm 8.4$ & $ 80.0\pm 9.0$  &            &                 &                 &   $  1.38$ & $1653.8\pm 5.6$ & $181.5\pm 4.9$  &    $ -2.48$ & $ 828.7\pm 6.9$ & $124.1\pm 8.0$ \\     
 $  7.54$ & $5454.0\pm 9.3$ & $ 88.9\pm 9.1$  &  \multicolumn{3}{c}{NGC 1308 - 5 arcsec E}     &   $  2.05$ & $1664.0\pm 5.5$ & $187.2\pm 5.0$  &    $ -2.20$ & $ 826.5\pm 6.5$ & $131.1\pm 7.7$ \\     
 $ 21.48$ & $5541.2\pm20.7$ & $ 72.6\pm16.2$  &            &                 &                 &   $  2.98$ & $1676.0\pm 5.5$ & $179.7\pm 4.9$  &    $ -1.93$ & $ 839.7\pm 6.7$ & $127.0\pm 7.7$ \\     
          &                 &                 &   $-19.45$ & $6176.9\pm20.0$ & $142.7\pm27.1$  &   $  4.19$ & $1667.9\pm 6.4$ & $179.2\pm 5.8$  &    $ -1.65$ & $ 835.3\pm 6.9$ & $133.2\pm 8.3$ \\   
\multicolumn{3}{c}{ESO 139-G009 - major axis} &   $ -8.25$ & $6257.5\pm10.0$ & $125.6\pm 9.2$  &   $  5.90$ & $1655.9\pm 5.9$ & $162.5\pm 5.3$  &    $ -1.38$ & $ 845.0\pm 5.9$ & $124.4\pm 6.8$ \\   
          &                 &                 &   $ -3.70$ & $6339.2\pm 8.6$ & $157.0\pm 9.1$  &   $  8.72$ & $1673.0\pm 7.6$ & $147.8\pm 6.2$  &    $ -1.10$ & $ 845.8\pm 5.4$ & $123.1\pm 6.2$ \\     
 $-24.33$ & $5174.2\pm 8.4$ & $ 64.6\pm 8.2$  &   $ -0.54$ & $6424.0\pm 7.7$ & $140.0\pm10.8$  &   $ 15.18$ & $1715.7\pm 7.5$ & $123.3\pm 5.7$  &    $ -0.83$ & $ 849.5\pm 5.2$ & $118.9\pm 6.0$ \\     
 $ -7.87$ & $5221.3\pm 9.2$ & $125.5\pm 8.8$  &   $  5.79$ & $6494.4\pm 9.3$ & $ 87.6\pm 9.2$  &   $ 32.73$ & $1748.6\pm 5.4$ & $ 59.8\pm 6.5$  &    $ -0.55$ & $ 850.8\pm 5.6$ & $117.0\pm 6.2$ \\     
 $ -3.22$ & $5285.5\pm15.6$ & $168.4\pm30.8$  &   $ 14.97$ & $6472.7\pm17.8$ & $ 69.4\pm22.4$  &            &                 &                 &    $ -0.28$ & $ 845.3\pm 5.5$ & $114.2\pm 6.1$ \\     
 $ -1.69$ & $5316.1\pm14.0$ & $180.5\pm31.6$  &            &                 &                 &  \multicolumn{3}{c}{NGC 1440 - 6 arcsec W}     &    $  0.00$ & $ 848.8\pm 5.2$ & $109.6\pm 5.6$ \\     
 $ -0.93$ & $5339.8\pm 7.0$ & $186.1\pm 5.7$  &  \multicolumn{3}{c}{NGC 1308 - major axis}     &            &                 &                 &    $  0.27$ & $ 840.2\pm 6.1$ & $111.4\pm 6.5$ \\     
 $ -0.39$ & $5355.3\pm 5.9$ & $202.4\pm 4.5$  &            &                 &                 &   $-42.55$ & $1463.1\pm13.3$ & $104.4\pm13.4$  &    $  0.55$ & $ 845.1\pm 5.2$ & $117.1\pm 5.7$ \\     
 $  0.14$ & $5382.3\pm 5.3$ & $190.0\pm 4.1$  &   $-15.22$ & $6131.4\pm12.6$ & $ 76.9\pm18.7$  &   $-25.35$ & $1458.2\pm 3.8$ & $ 67.8\pm 4.8$  &    $  0.82$ & $ 849.7\pm 6.6$ & $124.5\pm 7.5$ \\     
 $  0.67$ & $5412.7\pm 5.5$ & $188.1\pm 4.1$  &   $ -3.69$ & $6244.0\pm10.2$ & $188.7\pm15.8$  &   $-14.37$ & $1498.4\pm 4.2$ & $125.4\pm 3.4$  &    $  1.10$ & $ 849.4\pm 5.7$ & $115.4\pm 6.1$ \\     
 $  1.32$ & $5429.6\pm 6.4$ & $171.6\pm 5.4$  &   $ -1.28$ & $6277.2\pm 8.1$ & $208.7\pm10.5$  &   $ -9.42$ & $1545.8\pm 4.4$ & $135.2\pm 3.4$  &    $  1.37$ & $ 854.4\pm 6.8$ & $127.9\pm 7.6$ \\     
 $  2.45$ & $5460.2\pm18.0$ & $173.3\pm25.6$  &   $ -0.28$ & $6321.6\pm 7.2$ & $218.9\pm10.9$  &   $ -6.40$ & $1568.3\pm 4.9$ & $155.7\pm 4.2$  &    $  1.65$ & $ 849.5\pm 7.0$ & $123.6\pm 7.3$ \\     
 $  5.24$ & $5505.2\pm 7.4$ & $143.6\pm 6.3$  &   $  0.48$ & $6368.7\pm 8.6$ & $231.8\pm11.7$  &   $ -4.27$ & $1580.3\pm 5.3$ & $167.9\pm 4.6$  &    $  1.92$ & $ 855.9\pm 7.9$ & $139.6\pm 9.2$ \\     
 $ 18.53$ & $5572.6\pm 8.3$ & $ 65.3\pm 7.9$  &   $  1.60$ & $6392.5\pm 9.5$ & $222.5\pm14.1$  &   $ -2.40$ & $1595.6\pm 4.4$ & $156.4\pm 3.7$  &    $  2.33$ & $ 857.8\pm 5.3$ & $121.7\pm 6.2$ \\     
          &                 &                 &   $  6.25$ & $6453.5\pm13.3$ & $158.4\pm14.8$  &   $ -0.54$ & $1615.8\pm 5.1$ & $170.8\pm 4.4$  &    $  2.88$ & $ 860.4\pm 5.8$ & $129.7\pm 7.2$ \\                   
\multicolumn{3}{c}{ESO 139-G009 - 8 arcsec S} &   $ 21.31$ & $6505.5\pm19.7$ & $ 63.4\pm19.0$  &   $  1.46$ & $1635.7\pm 5.0$ & $164.6\pm 4.2$  &    $  3.43$ & $ 872.5\pm 6.3$ & $129.1\pm 7.3$ \\     
          &                 &                 &            &                 &                 &   $  3.94$ & $1643.5\pm 4.7$ & $141.7\pm 3.6$  &    $  4.11$ & $ 881.5\pm 6.1$ & $128.6\pm 7.2$ \\     
 $-19.38$ & $5217.0\pm15.6$ & $ 74.9\pm18.4$  &  \multicolumn{3}{c}{NGC 1308 - 5 arcsec W}     &   $  8.52$ & $1683.5\pm 6.5$ & $137.0\pm 5.8$  &    $  4.94$ & $ 885.9\pm 7.2$ & $114.6\pm 8.1$ \\     
 $ -5.66$ & $5279.9\pm 7.7$ & $ 83.6\pm 7.4$  &            &                 &                 &   $ 22.89$ & $1744.0\pm 4.5$ & $ 73.9\pm 6.6$  &    $  5.89$ & $ 893.3\pm 6.5$ & $105.4\pm 6.7$ \\     
 $  1.05$ & $5391.0\pm10.2$ & $102.0\pm10.6$  &   $-12.88$ & $6124.4\pm11.3$ & $ 67.1\pm13.5$  &   $ 41.65$ & $1748.0\pm 9.4$ & $ 85.0\pm 9.6$  &    $  7.12$ & $ 895.8\pm 6.2$ & $ 99.0\pm 6.1$ \\     
 $ 14.92$ & $5473.6\pm25.4$ & $132.4\pm23.6$  &   $ -2.74$ & $6173.8\pm 9.5$ & $122.8\pm10.2$  &            &                 &                 &    $  8.75$ & $ 909.4\pm 8.4$ & $102.2\pm 8.8$ \\     
          &                 &                 &   $  1.57$ & $6277.5\pm 9.2$ & $148.0\pm12.6$  &  \multicolumn{3}{c}{NGC 1440 - 11 arcsec W}    &    $ 11.15$ & $ 922.1\pm 8.3$ & $101.3\pm 8.3$ \\              
\multicolumn{3}{c}{IC 874 - 10 arcsec E}      &   $  4.39$ & $6330.4\pm 8.1$ & $141.0\pm10.9$  &            &                 &                 &    $ 14.94$ & $ 933.1\pm 9.3$ & $101.8\pm 9.6$ \\     
          &                 &                 &   $  8.92$ & $6403.7\pm12.7$ & $164.1\pm17.2$  &   $-36.82$ & $1461.5\pm 7.5$ & $ 75.1\pm 9.0$  &    $ 20.74$ & $ 949.1\pm11.5$ & $ 99.5\pm12.7$ \\     
 $-33.45$ & $2247.4\pm12.4$ & $ 89.1\pm19.0$  &   $ 19.76$ & $6481.6\pm13.7$ & $ 84.2\pm10.5$  &   $-19.80$ & $1498.6\pm 4.5$ & $ 86.2\pm 5.3$  &    $ 28.62$ & $ 948.8\pm13.1$ & $ 95.5\pm15.0$ \\     
 $-16.40$ & $2239.7\pm 8.2$ & $ 71.4\pm 8.3$  &            &                 &                 &   $-11.21$ & $1549.9\pm 5.8$ & $118.5\pm 4.4$  &    $ 39.03$ & $ 955.0\pm13.4$ & $ 81.5\pm14.4$ \\     
 $ -7.69$ & $2272.4\pm 4.4$ & $ 68.0\pm 4.3$  &  \multicolumn{3}{c}{NGC 1440 - 9 arcsec E}     &   $ -3.37$ & $1592.1\pm 7.1$ & $137.4\pm 5.7$  &    $ 52.46$ & $ 957.9\pm16.3$ & $ 92.8\pm21.1$ \\     
 $ -3.89$ & $2293.8\pm 4.4$ & $ 67.3\pm 3.9$  &            &                 &                 &   $  8.03$ & $1688.1\pm 6.4$ & $105.5\pm 6.7$  &    $ 69.24$ & $ 949.7\pm33.3$ & $ 72.6\pm51.3$ \\     
 $  1.57$ & $2345.2\pm 6.8$ & $ 79.4\pm 6.1$  &   $-28.77$ & $1471.1\pm 6.5$ & $ 80.6\pm 7.0$  &   $ 28.29$ & $1729.6\pm 7.1$ & $ 71.9\pm 8.3$  &             &                 &                \\     
 $ 15.11$ & $2376.9\pm 8.8$ & $ 54.0\pm 6.7$  &   $-12.03$ & $1503.8\pm 5.8$ & $109.0\pm 5.7$  &            &                 &                 &   \multicolumn{3}{c}{NGC 3412 - 10 arcsec W}   \\     
          &                 &                 &   $ -1.21$ & $1586.3\pm 4.6$ & $125.2\pm 3.4$  &  \multicolumn{3}{c}{NGC 3412 - 10 arcsec E}    &             &                 &                \\        
\multicolumn{3}{c}{IC 874 - major axis}       &   $  5.53$ & $1631.6\pm 4.5$ & $118.3\pm 3.7$  &            &                 &                 &    $-57.08$ & $ 717.2\pm17.6$ & $ 98.8\pm20.4$ \\     
          &                 &                 &   $ 12.49$ & $1680.8\pm 3.3$ & $ 95.9\pm 3.6$  &   $-67.31$ & $ 701.8\pm24.9$ & $114.3\pm35.1$  &    $-39.24$ & $ 744.2\pm11.7$ & $ 89.0\pm12.7$ \\     
 $-28.36$ & $2214.2\pm 9.1$ & $ 43.3\pm13.9$  &   $ 23.74$ & $1732.9\pm 3.7$ & $ 70.3\pm 5.0$  &   $-48.21$ & $ 729.4\pm12.6$ & $ 91.4\pm13.5$  &    $-27.61$ & $ 765.3\pm 8.5$ & $ 82.4\pm 6.9$ \\     
 $-11.17$ & $2213.6\pm 8.3$ & $ 86.0\pm10.8$  &   $ 38.51$ & $1764.4\pm 6.6$ & $ 62.0\pm 9.2$  &   $-34.56$ & $ 727.8\pm12.4$ & $112.2\pm14.4$  &    $-19.30$ & $ 779.4\pm 7.3$ & $ 86.5\pm 6.4$ \\     
 $ -6.73$ & $2238.7\pm 5.9$ & $ 96.2\pm 7.0$  &            &                 &                 &   $-24.17$ & $ 760.9\pm 9.2$ & $ 93.2\pm 9.0$  &    $-13.62$ & $ 801.1\pm 7.0$ & $ 89.8\pm 6.3$ \\     
 $ -4.72$ & $2252.2\pm 3.9$ & $ 93.8\pm 4.7$  &  \multicolumn{3}{c}{NGC 1440 - 6 arcsec E}     &   $-15.93$ & $ 761.9\pm 9.3$ & $103.7\pm 9.8$  &    $ -9.68$ & $ 815.3\pm 5.6$ & $ 78.4\pm 4.4$ \\     
 $ -3.29$ & $2272.8\pm 3.6$ & $ 99.6\pm 4.4$  &            &                 &                 &   $ -9.77$ & $ 793.5\pm 8.6$ & $105.9\pm 9.3$  &    $ -6.71$ & $ 831.0\pm 6.1$ & $ 96.1\pm 5.9$ \\     
 $ -2.35$ & $2267.4\pm 3.4$ & $ 98.0\pm 4.2$  &   $-39.57$ & $1443.5\pm 8.3$ & $ 76.6\pm 9.5$  &   $ -5.15$ & $ 816.4\pm 6.4$ & $100.7\pm 6.5$  &    $ -3.98$ & $ 848.6\pm 6.7$ & $103.0\pm 6.9$ \\     
 $ -1.69$ & $2262.6\pm 3.1$ & $ 96.6\pm 3.9$  &   $-21.32$ & $1458.5\pm 5.2$ & $ 90.4\pm 5.7$  &   $ -1.47$ & $ 827.6\pm 7.5$ & $109.1\pm 8.2$  &    $ -1.25$ & $ 860.7\pm 6.3$ & $ 95.1\pm 6.1$ \\     
 $ -1.15$ & $2276.7\pm 2.3$ & $101.4\pm 3.0$  &   $ -8.82$ & $1528.2\pm 4.8$ & $136.8\pm 3.8$  &   $  1.67$ & $ 844.8\pm 6.7$ & $101.7\pm 6.8$  &    $  1.62$ & $ 873.5\pm 6.4$ & $ 93.7\pm 6.0$ \\     
 $ -0.76$ & $2287.6\pm 2.4$ & $101.1\pm 2.9$  &   $ -4.42$ & $1558.9\pm 4.4$ & $142.6\pm 3.4$  &   $  4.54$ & $ 865.3\pm 6.5$ & $105.4\pm 7.1$  &    $  4.87$ & $ 882.8\pm 7.2$ & $ 92.4\pm 6.6$ \\     
 $ -0.49$ & $2287.4\pm 2.5$ & $106.2\pm 3.2$  &   $ -1.94$ & $1564.0\pm 4.5$ & $149.5\pm 3.9$  &   $  7.40$ & $ 876.0\pm 6.9$ & $ 93.0\pm 6.3$  &    $  8.92$ & $ 902.7\pm 8.1$ & $ 99.1\pm 8.4$ \\     
 $ -0.22$ & $2299.4\pm 2.0$ & $ 97.4\pm 2.5$  &   $  0.06$ & $1596.3\pm 4.5$ & $152.8\pm 3.6$  &   $ 10.63$ & $ 897.2\pm 8.3$ & $ 97.9\pm 8.3$  &    $ 14.56$ & $ 938.0\pm 7.4$ & $ 85.6\pm 6.0$ \\     
 $  0.05$ & $2312.4\pm 2.1$ & $100.7\pm 2.8$  &   $  1.92$ & $1607.7\pm 4.3$ & $161.9\pm 3.7$  &   $ 14.84$ & $ 907.4\pm 7.3$ & $ 85.4\pm 6.3$  &    $ 22.24$ & $ 944.8\pm10.1$ & $ 98.7\pm10.9$ \\     
 $  0.32$ & $2317.2\pm 2.1$ & $ 98.5\pm 2.6$  &   $  3.92$ & $1625.2\pm 4.5$ & $153.1\pm 4.0$  &   $ 20.82$ & $ 919.4\pm10.3$ & $108.9\pm11.5$  &    $ 32.26$ & $ 957.8\pm11.8$ & $ 96.1\pm13.8$ \\     
 $  0.59$ & $2329.3\pm 2.4$ & $104.6\pm 3.1$  &   $  6.32$ & $1636.3\pm 4.2$ & $134.3\pm 3.3$  &   $ 29.48$ & $ 935.7\pm14.0$ & $108.6\pm17.9$  &    $ 45.31$ & $ 978.9\pm15.3$ & $106.7\pm19.5$ \\     
 $  0.86$ & $2340.2\pm 2.9$ & $108.0\pm 3.9$  &   $  9.60$ & $1659.9\pm 4.4$ & $132.9\pm 3.6$  &   $ 41.32$ & $ 962.1\pm18.8$ & $132.3\pm25.4$  &    $ 63.88$ & $ 955.0\pm18.6$ & $ 69.6\pm27.0$ \\     
 $  1.25$ & $2353.0\pm 2.2$ & $ 95.9\pm 2.9$  &   $ 15.16$ & $1686.4\pm 5.2$ & $119.3\pm 4.6$  &   $ 59.71$ & $ 964.5\pm16.9$ & $ 97.4\pm22.3$  &             &                 &                \\     
 $  1.79$ & $2356.2\pm 3.0$ & $106.2\pm 4.1$  &   $ 30.70$ & $1755.2\pm 4.3$ & $ 71.1\pm 5.8$  &            &                 &                 &   \multicolumn{3}{c}{NGC 3412 - 15 arcsec W}   \\     
 $  2.45$ & $2355.7\pm 3.2$ & $103.8\pm 4.1$  &            &                 &                 &  \multicolumn{3}{c}{NGC 3412 - major axis}     &             &                 &                \\     
 $  3.51$ & $2351.7\pm 3.2$ & $ 94.9\pm 3.6$  &  \multicolumn{3}{c}{NGC 1440 - major axis}     &            &                 &                 &    $-61.49$ & $ 725.2\pm29.0$ & $108.3\pm38.2$ \\     
 $  5.20$ & $2356.5\pm 3.5$ & $ 80.0\pm 3.1$  &            &                 &                 &   $-70.26$ & $ 720.5\pm33.0$ & $ 82.9\pm51.5$  &    $-39.65$ & $ 741.8\pm13.9$ & $ 77.4\pm15.5$ \\   
 $  7.03$ & $2385.7\pm 8.1$ & $ 95.3\pm 9.5$  &   $-40.26$ & $1444.5\pm 7.4$ & $ 40.4\pm 7.9$  &   $-52.52$ & $ 751.3\pm17.1$ & $ 96.2\pm20.9$  &    $-25.09$ & $ 755.6\pm 9.9$ & $ 95.9\pm 9.7$ \\     
 $ 13.29$ & $2411.7\pm 7.4$ & $ 71.1\pm 6.1$  &   $-23.27$ & $1455.5\pm 4.6$ & $ 74.8\pm 6.0$  &   $-38.00$ & $ 756.9\pm13.3$ & $100.7\pm15.3$  &    $-14.98$ & $ 800.7\pm 8.2$ & $ 83.8\pm 7.2$ \\     
 $ 32.84$ & $2397.4\pm 8.0$ & $ 63.6\pm14.6$  &   $-12.34$ & $1509.3\pm 5.8$ & $121.0\pm 4.6$  &   $-27.50$ & $ 759.4\pm10.8$ & $ 87.9\pm10.8$  &    $ -8.14$ & $ 824.5\pm 6.9$ & $ 80.9\pm 5.4$ \\     
          &                 &                 &   $ -8.05$ & $1544.7\pm 5.4$ & $140.4\pm 4.3$  &   $-19.78$ & $ 755.6\pm10.3$ & $ 98.4\pm10.6$  &    $ -2.65$ & $ 840.0\pm 8.0$ & $ 95.7\pm 7.9$ \\     
\multicolumn{3}{c}{IC 874 - 10 arcsec W}      &   $ -5.70$ & $1541.1\pm 5.2$ & $159.4\pm 4.3$  &   $-14.24$ & $ 777.7\pm 7.5$ & $ 92.8\pm 6.7$  &    $  3.32$ & $ 866.8\pm 8.4$ & $ 92.8\pm 8.1$ \\     
          &                 &                 &   $ -4.10$ & $1537.0\pm 6.0$ & $171.9\pm 5.3$  &   $-10.77$ & $ 796.7\pm 7.4$ & $105.9\pm 7.9$  &    $ 10.81$ & $ 910.3\pm10.2$ & $104.4\pm10.9$ \\     
 $-30.18$ & $2243.4\pm17.0$ & $ 79.7\pm12.5$  &   $ -2.90$ & $1542.6\pm 5.4$ & $182.0\pm 4.9$  &   $ -8.47$ & $ 810.5\pm 6.8$ & $107.7\pm 7.5$  &    $ 20.71$ & $ 923.8\pm10.7$ & $ 89.1\pm10.9$ \\     
 $-12.54$ & $2266.0\pm 6.8$ & $ 65.1\pm 6.2$  &   $ -1.96$ & $1547.9\pm 5.7$ & $195.5\pm 5.6$  &   $ -6.84$ & $ 804.4\pm 6.9$ & $110.4\pm 7.3$  &    $ 33.33$ & $ 945.6\pm14.5$ & $ 87.1\pm17.0$ \\   
 $ -1.34$ & $2296.5\pm 5.6$ & $ 67.1\pm 4.9$  &   $ -1.30$ & $1547.8\pm 5.9$ & $191.7\pm 6.0$  &   $ -5.61$ & $ 808.9\pm 5.7$ & $107.7\pm 6.0$  &    $ 52.54$ & $ 952.0\pm15.5$ & $ 76.4\pm18.6$ \\   
 $  3.54$ & $2335.1\pm 4.4$ & $ 69.4\pm 4.0$  &   $ -0.76$ & $1567.3\pm 5.4$ & $191.1\pm 5.3$  &   $ -4.66$ & $ 803.5\pm 5.2$ & $106.5\pm 5.4$  &    $      $   $             $   $            $ \\                                   
\hline 
\label{tab:kinematics}
\end{tabular}   
\end{tiny}
\end{table}    
\end{landscape}

\subsection{Pattern speed measurement} 

We measured the bar pattern speed for each sample galaxy by applying
the TW method as done in Paper I.

To compute the mean position of stars, $\pin$, along the slits, we
generally extracted profiles from $I$ and $V$-band surface photometry
along the positions of the slits.  In the case of NGC 1440, because of
the radial colour gradient, we preferred to use the intensity from the
slit spectra themselves, although they are somewhat noisier.  For the
remaining galaxies, the $I$-band profiles match very well the profiles
obtained by collapsing the spectra along the wavelength direction,
confirming that the slits were placed as intended.  We used the broad
band profiles to compute $\pin$ because these are less noisy than the
spectral profiles, particularly at large radii.  We computed the value
of $\pin$ at each slit position by Monte Carlo simulation, with
photon, read-out and sky noise to compute the errors.  Formally, the
integrals in TW equation are over $-\infty \leq X \leq \infty$, but
can be limited to $-X_{max} \leq X \leq X_{max}$ if $X_{max}$ has
reached the axisymmetric part of the disc; still larger values of
$X_{max}$ add noise only.  The $\pin$ values thus obtained are given
in Tab.  \ref{tab:TW_values}.

To measure the luminosity-weighted line-of-sight stellar velocity,
$\kin$, for each slit position, we collapsed each two-dimensional
spectrum along its spatial direction to obtain a one-dimensional
spectrum.
The resulting spectra have been analysed with the FCQ method using the
same template star adopted in Sect. \ref{sec:kinematics} to derive the
stellar kinematics of the galaxy; $\kin$ is then the radial velocity
derived from the LOSVD of the one-dimensional spectra.  For each slit
position, the uncertainty on $\kin$ was estimated by means of Monte
Carlo simulations as done for $v$ in measuring the stellar kinematics.
The $\kin$ values we derived along each slit and the adopted
wavelength (always including the Mg triplet) and radial ranges
(limited by the noise, and after removing the contribution of
foreground stars by linear interpolation) are given in Tab.
\ref{tab:TW_values}.

\begin{figure*}
 \leavevmode{\psfig{figure=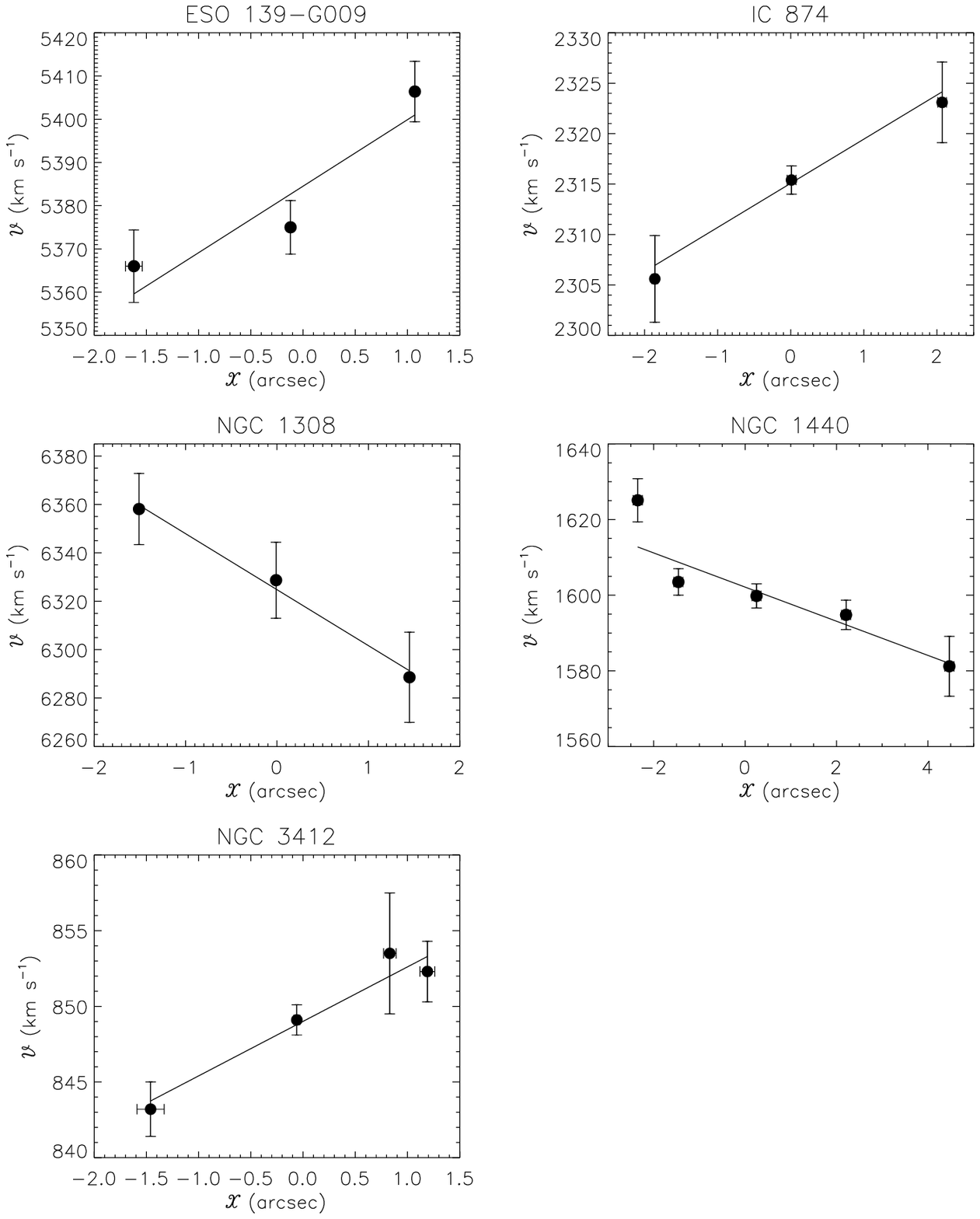,height=22.truecm,width=15.truecm}}
\caption[]{Pattern speed measurements for our sample of galaxies.  The
  straight line is the best fit, with slope $\om \sin i$.  The
  resulting values of $\om$ are given in Table
  \ref{tab:bar_kinematics}}
\label{fig:corotations}
\end{figure*}

For each galaxy, we derived $\om \sin i$ by fitting a straight line to
the values $(\pin,\kin)$ obtained from the available slit positions
(Fig.  \ref{fig:corotations}).  Finally the value of $\om$ (Table
\ref{tab:bar_kinematics}) was derived from the adopted galaxy
inclination, as given in Table \ref{tab:measured_pa}.

\begin{table*}
\centering
\caption{The values of  $\kin$ and $\pin$ for each  slit of our sample
of galaxies.}
\begin{tabular}{llrrcc} 
\hline
\multicolumn{1}{c}{Galaxy}  & 
\multicolumn{1}{c}{Offset} &  
\multicolumn{1}{c}{$\pin$} &  
\multicolumn{1}{c}{$\kin$} &  
\multicolumn{1}{c}{$|X|_{{\rm max}}$} &  
\multicolumn{1}{c}{$\lambda_{{\rm min}},\lambda_{{\rm max}}$} \\
\multicolumn{1}{c}{} &   
\multicolumn{1}{c}{} &  
\multicolumn{1}{c}{(arcsec)} &  
\multicolumn{1}{c}{(\kms)} &  
\multicolumn{1}{c}{(arcsec)} &
\multicolumn{1}{c}{(\AA)}  \\   
\hline
ESO 139-G009 & $10$ arcsec N & $+1.07 \pm 0.03$ & $5406.4 \pm  7.0$ & 
 $40$ & 5014.1,5486.2\\ 
          & major axis    & $-0.12 \pm 0.02$ & $5375.0 \pm  6.2$ & 
 $60$ & 5014.1,5486.2\\
          & $ 8$ arcsec S & $-1.62 \pm 0.08$ & $5366.0 \pm  8.4$ & 
 $50$ & 5014.1,5486.2\\ 
\hline                                                              
IC 874    & $10$ arcsec E & $-1.86 \pm 0.04$ & $2305.6 \pm  4.3$ & 
 $55$ & 5039.2,5486.2\\ 
          & major axis    & $+0.01 \pm 0.06$ & $2315.4 \pm  1.4$ & 
 $80$ & 5039.2,5486.2\\ 
          & $10$ arcsec W & $+2.07 \pm 0.06$ & $2323.1 \pm  4.0$ & 
 $55$ & 5039.2,5486.2\\ 
\hline                                                              
NGC 1308  & $ 5$ arcsec E & $-1.51 \pm 0.01$ & $6358.1 \pm 14.7$ & 
 $35$ & 5115.3,5431.7\\ 
          & major axis    & $-0.01 \pm 0.00$ & $6328.7 \pm 15.7$ & 
 $35$ & 5115.3,5431.7\\ 
          & $ 5$ arcsec W & $+1.45 \pm 0.01$ & $6288.6 \pm 18.7$ & 
 $35$ & 5115.3,5431.7\\ 
\hline                                                              
NGC 1440  & $ 9$ arcsec E & $-2.35 \pm 0.10$ & $1625.1 \pm  5.7$ & 
 $50$ & 5115.3,5431.7\\ 
          & $ 6$ arcsec E & $-1.46 \pm 0.10$ & $1603.5 \pm  3.5$ & 
 $50$ & 5115.3,5431.7\\ 
          & major axis    & $+0.25 \pm 0.10$ & $1599.8 \pm  3.2$ & 
 $50$ & 5115.3,5431.7\\ 
          & $ 6$ arcsec W & $+2.21 \pm 0.10$ & $1594.8 \pm  3.9$ & 
 $50$ & 5115.3,5431.7\\ 
          & $11$ arcsec W & $+4.47 \pm 0.10$ & $1581.2 \pm  7.9$ & 
 $50$ & 5115.3,5431.7\\ 
\hline                                                              
NGC 3412  & $10$ arcsec E & $-1.46 \pm 0.13$ & $ 843.2 \pm  1.8$ & 
 $80$ & 4964.2,5541.4\\ 
          & major axis    & $-0.06 \pm 0.03$ & $ 849.1 \pm  1.0$ & 
 $80$ & 4964.2,5541.4\\ 
          & $10$ arcsec W & $+1.19 \pm 0.07$ & $ 852.3 \pm  2.0$ & 
 $80$ & 4964.2,5541.4\\ 
          & $15$ arcsec W & $+0.83 \pm 0.06$ & $ 853.5 \pm  4.0$ & 
 $80$ & 4964.2,5541.4\\ 
\hline
\end{tabular}
\label{tab:TW_values}
\end{table*}

\subsection{Systemic and mean streaming velocities}

To measure the systemic velocity of the galaxies, we fit interpolating
splines to the measured velocities and fitted ``tilted rings'', of
fixed $i$ and PA (as obtained from the photometry), to the inner parts
of the galaxies.  In the case of NGC 1308, our data were not of
sufficient S/N to permit such fits, in which case we used $\kin$ on
the major axis; for the other four galaxies, comparision of Tables
\ref{tab:TW_values} and \ref{tab:bar_kinematics} shows that the
major-axis $\kin$ is a very good approximation to $V_{\rm sys}$.  In
Table \ref{tab:bar_kinematics}, we compare our measurements of $V_{\rm
  sys}$ with those in RC3.  In all cases we find excellent agreement.

Once $V_{\rm sys}$ is measured, we obtain the stellar mean streaming
velocities, $V_*$, by subtracting $V_{\rm sys}$, then folding about
the origin the major-axis data.  Figure \ref{fig:asym_drift_corr}
shows that our folded spectra have no substantial asymmetries on the
two sides.

\subsection{Asymmetric drift correction}

Measurement of $\om$ with the TW method requires no modeling.
However, in the absence of gas velocities at large radii, measurement
of $\vpd$ requires some modeling to recover the rotation curve from
the observed stellar streaming velocities.  This asymmetric drift
correction can be fairly large in early-type disc galaxies, where the
velocity dispersions are large.  We start from the asymmetric drift
equation (\eg\ Binney \& Tremaine [1987] Eqn. 4-33)
\begin{equation}
V_c^2 - V_*^2 =  -\sigma_R^2 \left[ {\frac{\partial \ln \rho}{\partial
\ln R}} + {\frac{\partial \ln  \sigma_R^2}{\partial \ln R}} + \left( 1
- {\frac{\sigma_\phi^2}{\sigma_R^2}} \right) \right],
\label{eqn:asymdrif}
\end{equation}
where $V_c$ is the circular velocity, $\rho$ is the disk's volume
density, and $\sigma_\phi$ and $\sigma_R$ are the tangential and
radial velocity dispersions in the cylindrical coordinates of the
galaxy's intrinsic plane.  We then make the following assumptions:

\begin{enumerate}
\item We assume that the true rotation curve is flat at large radii.
  This is usually a very good approximation in high surface-brightness
  galaxies.  Thus we only measure $V_{\rm c,flat}$, the amplitude of
  the rotation curve.  A flat rotation curve implies
  $\sigma_\phi^2/\sigma_R^2 = \frac{1}{2}$, which allows us to write,
  for the observed velocity dispersion on the major-axis, $\sigma_{\rm
    obs} = \sigma_\phi \sin i \sqrt{1 + 2 \alpha^2 \cot^2i}$, where
  $\alpha = \sigma_z/\sigma_R$.  We use 3 values of $\alpha$, 0.7,
  0.85 and 1.0, which bracket a plausible range for this parameter in
  early-type disc galaxies extrapolating from the only 3 measured
  values in later-type systems (Dehnen \& Binney 1998; Gerssen \etal\ 
  1997; Gerssen \etal\ 2000).  The variations in this parameter are
  included in our error estimate of $V_{\rm c,flat}$.
  
\item We assume that $\sigma_{\rm obs}$ decreases exponentially with
  radius, with scale-length $R_\sigma$.  We fit $R_\sigma$ with the
  major-axis data, outside the bar region when the S/N permits; these
  fits are shown in Fig. \ref{fig:asym_drift_corr}.

\item We assume the the disc's volume density is radially exponential.
  Since we are interested here in the total mass density, we fitted an
  exponential to the surface brightness measured in ellipses with
  position angle and ellipticity fixed to the values determined from
  large radii.  The fits to $R_{\rm d}$ started outside $\len$; we
  varied the radial range of the fits to obtain error estimates on
  $R_{\rm d}$; these were as large as 15 per cent.  Variations in
  $R_{\rm d}$ are also included in our error estimate on $V_{\rm c,flat}$.
\end{enumerate}

With these assumptions, the asymmetric drift equation becomes
\begin{equation}
V_c^2 =  V_*^2 +  \frac{\sigma_{\rm obs}^2}{\sin^2 i  (1 +  2 \alpha^2
\cot^2i)} \left[  2 R \left( \frac{1}{R_{\rm  d}} + \frac{2}{R_\sigma}
\right) - 1 \right].
\end{equation}
We apply this correction to all velocity data points (including those
not on the major-axis) outside the bar radius and within $30\degrees$
of the major-axis.  We average over all these points to obtain $V_{\rm
  c,flat}$, with an additional error from the scatter of the points 
added in quadrature.  We report our findings of $V_{\rm c,flat}$ in Table
\ref{tab:bar_kinematics}, and in Fig. \ref{fig:asym_drift_corr}.

In Paper I, the high surface brightness of NGC 1023 allowed us to drop
the assumption that $\sigma_\phi^2/\sigma_R^2 = \frac{1}{2}$, and
instead we used the velocity dispersion data from all slits to obtain
$V_{\rm c,flat} = 270 \pm 31$ \kms\ (having assumed $0.0 \leq \alpha
\leq 1.0$; because of its high inclination $V_{\rm c,flat}$ for this
galaxy is not very sensitive to the value of $\alpha$).  For
comparison, we have redone the asymmetric drift correction of NGC 1023
with the method described above for the new sample.  We found $V_{\rm
  c,flat} = 270 \pm 13$ \kms, which is in excellent agreement with the
value of Paper I.

We finally measured $\vpd$ as $V_{\rm c,flat}/(\len\om)$.  We used
Monte Carlo to estimate the uncertainties in $\vpd$ by varying $V_{\rm
  c,flat}$ and $\len$ uniformly in their respective ranges, and
varying $\om$ assuming the errors of Table \ref{tab:bar_kinematics}
are Gaussian.  We report the median and the 67 per cent interval of
$\vpd$ in Table \ref{tab:bar_kinematics}.

\begin{figure}
\leavevmode{\psfig{figure=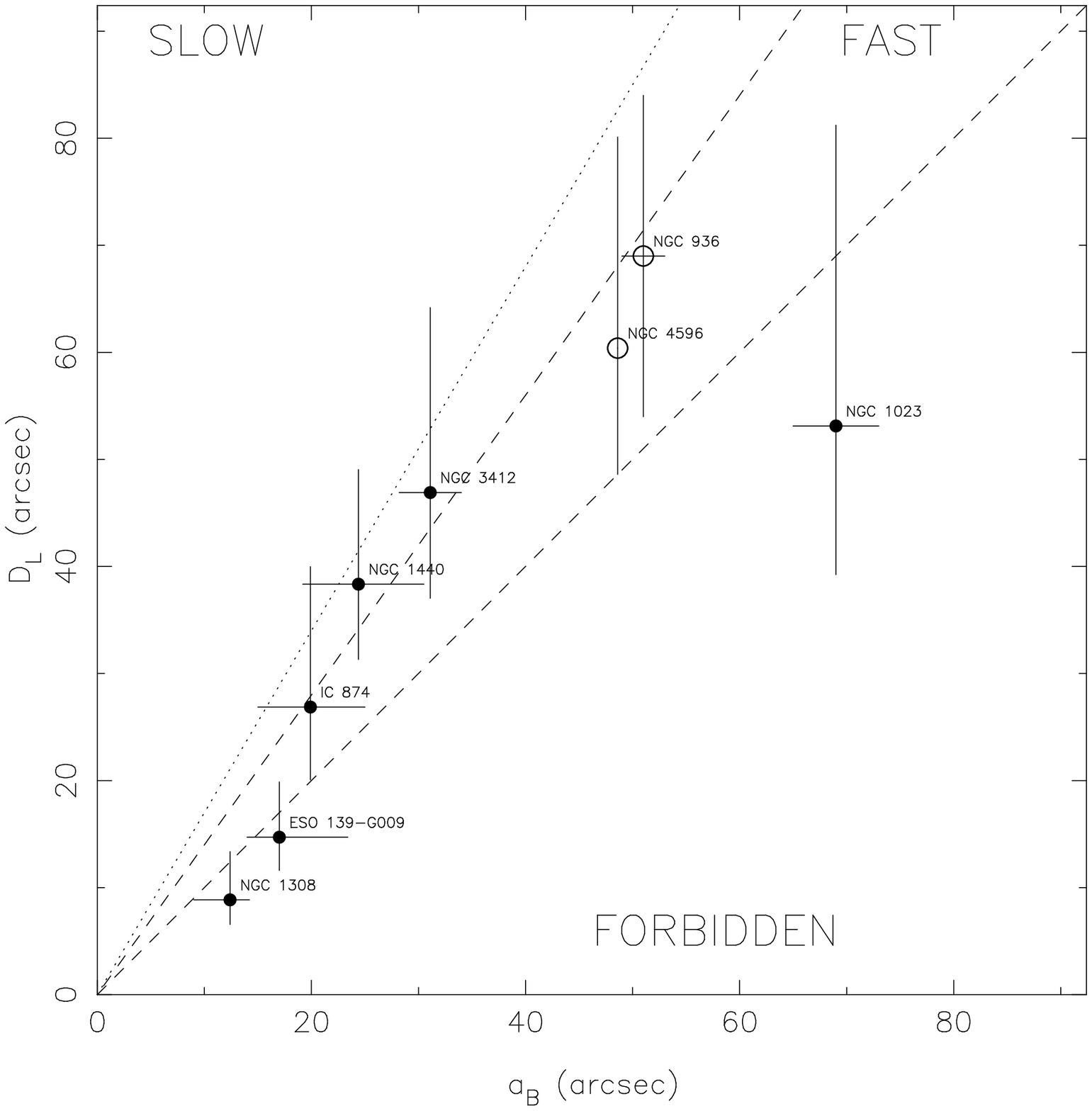,width=8.5truecm}}
\caption[]{The corotation radius, $D_{L}$, and the bar semi-major axis, 
  $a_{B}$, for the sample galaxies, including NGC 1023 (Paper I).  The
  open circles represent galaxies which are not part of our sample:
  NGC 936 (Merrifield \& Kuijken 1995) and NGC 4596 (Gerssen \etal\ 
  1999).  Dashed lines corresponding to $\vpd=1$ and $\vpd=1.4$,
  separate the fast-bar, slow-bar and forbidden regimes. The dotted
  line corresponds to $\vpd=1.7$.}
\label{fig:scatter}
\end{figure}

\begin{table*}   
\begin{center}   
\caption{Disc and bar kinematics}
\begin{tabular}{l r r r c c c}   
\hline \multicolumn{1}{c}{Galaxy}  & \multicolumn{1}{c}{$V_{\rm sys}$} &  
\multicolumn{1}{c}{$V_{\rm sys,RC3}$}  & 
\multicolumn{1}{c}{$V_{\rm c,flat}$} & \multicolumn{1}{c}{$\om$} &  
\multicolumn{1}{c}{$\lag$} & \multicolumn{1}{c}{$\vpd$} \\ 
&  (\kms) & (\kms) & (\kms)  & (\kmsa) & (arcsec)  & \\  
\hline 
ESO 139-G009 & $5377\pm 6$ & $5370\pm60$ & $314^{+14}_{-12}$ & 
 $21.4\pm 5.8$ & $14.7^{ +5.2}_{-3.1}$ & $0.8^{+0.3}_{-0.2}$\\  
IC 874       & $2309\pm 4$ & $2311\pm36$ & $187^{+14}_{-13}$ & 
 $ 7.0\pm 2.4$ & $26.9^{+13.1}_{-6.8}$ & $1.4^{+0.7}_{-0.4}$\\  
NGC 1308     & $6329\pm16$ & $6350\pm42$ & $347^{+21}_{-18}$ & 
 $39.7\pm13.9$ &  $8.9^{ +4.5}_{-2.3}$ & $0.8^{+0.4}_{-0.2}$\\  
NGC 1440     & $1601\pm 2$ & $1534\pm66$ & $283^{+11}_{-9}$ & 
 $ 7.4\pm 1.7$ & $38.3^{+10.7}_{-7.0}$ & $1.6^{+0.5}_{-0.3}$\\  
NGC 3412     &  $850\pm 2$ &  $865\pm27$ & $205^{+18}_{-15}$ & 
 $ 4.4\pm 1.2$ & $46.9^{+17.3}_{-9.9}$ & $1.5^{+0.6}_{-0.3}$\\ 
\hline
\label{tab:bar_kinematics}   
\end{tabular}   
\end{center}   
\end{table*}

\section{Discussion and Conclusions}
\label{sec:conclusions}

The 5 SB0's presented in this work, together with NGC 1023 studied in
Paper I, which we include in our sample for this discussion, represent the 
largest sample of barred galaxies, with $\om$ measured by means of the 
TW method.  For all of them, $\vpd$ is consistent with being in the range 
1.0 to 1.4, within the errors, \ie\ with each having a fast bar.  The 
unweighted average for the sample is $\overline{\vpd} = 1.1$.
The apparent range of $\vpd$ spans from 0.8 to 1.6 (0.6 to 2.1, within 
the 67 per cent intervals).  This spread 
is not related to the properties of the galaxies in any obvious way (\eg\
for the two galaxies at $V_{\rm c, flat} = 277$ \kms, NGC 1023 and NGC 
1440, the measured values of $\vpd$ are at opposite extremes of the 
distribution).  The fact that some of the values of $\vpd$ are nominally
less than unity leads us to suggest that the large range of $\vpd$ is a
result of random errors and/or scatter in the measurements.

The sources of random errors are largely due to measurement
uncertainties in all 3 quantities used to compute $\vpd$, \ie\ $V_{\rm
  c,flat}$, $\len$ and $\om$.  Of these, the largest is in $\om$,
amounting to typical fractional uncertainties of 30 per cent, followed
by $\len$, for which the typical fractional uncertainty is 20 per
cent.  These uncertainties account for the typical large (and
asymmetric towards large values) errors on measurements of $\vpd$.

A likely source of scatter is errors in the disc PA.  Debattista (2002) 
shows that, for PA errors of FWHM $5\degrees$ (note that the root-mean-square
PA uncertainty in our sample is $2\fdg1$), the scatter in $\vpd$ is 
of order 0.44, large enough to account for the $\vpd < 1$ cases.  Since
PA errors scatter $\vpd$ to both larger and smaller values, then the largest
measured value of $\vpd = 1.53$ is probably an over-estimate.
If this is the case, then TW measurements are finding the same range of 
$\vpd$ as do hydrodynamic simulations.

The conclusion that bars are fast constrains the dark matter
distribution in disk galaxies.  Debattista \& Sellwood (1998, 2000)
argued that bars this fast can only survive if the disc in which they
formed is maximal.  Recent high resolution $N$-body simulations with
cosmologically-motivated dark matter halos produce bars with $\vpd$ in
the range between 1.2 and 1.7 (Valenzuela \& Klypin 2002).  Even
discounting our argument above in favor of a more restricted range of
$\vpd$, Fig.  \ref{fig:scatter} shows that $\vpd = 1.7$ is possible
only for the bars of IC 874, NGC 1440 NGC 3412 and, marginally, NGC
936, while the bars of ESO 139-G009, NGC 1023, NGC 1308 and NGC 4596
never reach this value of $\vpd$.  Note, moreover, that 3 of the
galaxies that do reach $\vpd = 1.7$ have amongst the largest
fractional errors in $\vpd$.  Therefore we conclude that the $N$-body
models of Valenzuela \& Klypin (2002) probably produce slower bars
than the observed.

The galaxy ESO 139-G009 is classified as SAB (\ie\ weakly barred) in RC3; 
the fact that it hosts a fast bar suggests that weak
bars form via the same mechanism that forms the strong ones.  
Thus the hypothesis of Kormendy (1979),
that weak bars are the end state of slowed down fast bars, already
in question from $N$-body simulations (Debattista \& Sellwood 2000),
is also unsupported by the limited observational data.  Further 
measurements of $\vpd$ for weak bars would be of considerable interest.

\bigskip
\noindent
{\bf Acknowledgements.}

\noindent
VPD and JALA acknowledge support by the Schweizerischer Nationalfonds
through grant 20-64856.01. JALA was partially supported by Spanish DGC
(Grant AYA2001-3939). EMC acknowledges the Astronomisches Institut der
Universit\"at Basel for the hospitality while this paper was in
progress.
We are indebted to R.  Bender and R. Saglia for providing us with the
FCQ package which we used for measuring the stellar kinematics.  We
are also grateful to A. Pizzella for the images he acquired.  This
research has made use of the Lyon-Meudon Extragalactic Database (LEDA)
and of the NASA/IPAC Extragalactic Database (NED).
This paper is based on observations carried out with the New
Technology Telescope and the Danish 1.54-m Telescope (Prop.  No.
67.B-0230 and No.  68.B-329) at the Europen Southern Observatory, La
Silla (Chile), with the Italian Telescopio Nazionale Galileo (Prop.
AOT-3, 3-06-119) operated on the island of La Palma by the Centro
Galileo Galilei of the Consorzio Nazionale per l'Astronomia e
l'Astrofisica, and with the Jacobus Kapteyn Telescope operated by the
Isaac Newton group at La Palma island at the Spanish Observatorio del
Roque de los Muchachos of the Instituto de Astrof\'\i sica de
Canarias.

\end{document}